\documentstyle[psfig,12pt]{article}
\begin{document}
\title{
Dilepton Production from AGS to SPS Energies within a Relativistic
Transport Approach
\thanks{Work supported by BMBF and GSI Darmstadt.}}
\author{E. L. Bratkovskaya and W. Cassing
\\ Institut f\"ur Theoretische Physik,
Universit\"at Giessen\\D-35392 Giessen, Germany}
\date{}
\maketitle

\begin{abstract}
We present a nonperturbative dynamical study of $e^+e^-$ production in
proton-nucleus and nucleus-nucleus collisions from AGS to SPS energies
on the basis of the covariant transport approach HSD.  For p~+~Be
reactions the dilepton yield for invariant masses  $M \leq 1.4$~GeV is
found to be dominated by the decays of the $\eta, \rho, \omega$ and
$\Phi$ mesons at all energies from 10 -- 450~GeV.  For nucleus-nucleus
collisions, however, the dilepton yield shows an additional large
contribution from $\pi^+\pi^-$, $K^+K^-$ and $\pi \rho$ channels.
Systematic studies are presented for the 'free' meson mass scenario in
comparison to a 'dropping' meson mass scenario at finite baryon
density.  We find that for 'dropping' meson masses the invariant
dilepton mass range 0.35~GeV $\leq M \leq$ 0.65~GeV is increased in
comparison to the 'free' meson mass scenario and that the data of the
CERES-collaboration for nucleus-nucleus collisions can be described
much better within the 'dropping' mass scheme.  We study in detail the
contributions from the various dilepton channels as a function of the
transverse momentum and rapidity of the lepton pair as well as a
function of the charged particle multiplicity.  Furthermore, various
direct photon channels for S~+~Au at 200~GeV/u are computed and found
to be well below the upper bounds measured by the WA80-collaboration.
\end{abstract}

\vspace{1cm}


\newpage
\section{Introduction}

The question of chiral symmetry restoration at high baryon density is
of fundamental interest since a couple of years~\cite{brownrho,chiral},
but a clear experimental evidence has not been achieved, so far. The
enhancement of strangeness production as e.g. seen in the AGS data for
the $K^+/\pi^+$ ratio \cite{kplus} might be a signiture for such a
transition \cite{Ehehalt}, however, other hadronic scenarios can be
cooked up to describe this phenomenon as well~\cite{Braun}.  Enhanced
antikaon yields at SIS energies, as compared to transport studies using
bare kaon masses~\cite{Fang,Cass96k}, point in the same direction, but
here the present knowledge on the elementary production cross sections
close to threshold does not yet allow for a final identification. On
the other hand, dileptons are particularly well suited for an
investigation of the violent phases of a high-energy heavy-ion
collision because they can leave the reaction volume essentially
undistorted by final-state interactions.  Indeed, dileptons from
heavy-ion collisions  have been observed by the DLS collaboration at
the BEVALAC \cite{ro88,na89,ro89} and by the CERES~\cite{CERES},
HELIOS~\cite{HELIOS,HELI2}, NA38~\cite{NA38} and NA50
collaborations~\cite{NA50} at SPS energies.

Quite some years ago it has been found within microscopic transport
studies at BEVALAC/SIS energies \cite{Wolf1,Ko90} that above about
0.5~GeV of invariant mass (of the lepton pair) the dominant production
channel is from $\pi^+ \pi^-$ annihilation, such that the properties of
the short lived $\rho$ meson could be explored at high baryon density.
The data available so far, however, did not allow for a closer
distinction of the various models proposed.

The recent data on $e^+e^-$ or $\mu^+ \mu^-$ spectra at SPS energies,
on the other hand, are more conclusive. The enhancement of the low mass
dimuon yield in S~+~W compared to p~+~W collisions \cite{HELIOS} at
200~GeV/u has been first suggested by Koch et al.~\cite{Koch} to be due
to $\pi^+\pi^-$ annihilation.  Furthermore, Li et al.~\cite{Li} have
proposed that the enhancement of the $e^+e^-$ yield in S~+~Au collision
as observed by the CERES collaboration~\cite{CERES} should be due to an
enhanced $\rho$-meson production (via $\pi^+\pi^-$ annihilation) and a
dropping $\rho$-mass in the medium. In fact, their analysis -- which
was based on an expanding fireball scenario in chemical equilibrium --
could be confirmed within the microscopic transport calculations in
Ref.~\cite{Cass95C}.  Meanwhile, various authors have substantiated the
observation in Refs.~\cite{Li,Cass95C}, that the spectral shape of the
dilepton yield is incompatible with 'free' meson form factors
\cite{Koch96,Srivas,Frankfurt}.  However, a more conventional approach
including the change of the $\rho$-meson spectral function in the
medium due to the coupling of the $\rho, \pi, \Delta$ and nucleon
dynamics along the line of Refs.~\cite{Herrmann,asakawa,Chanfray} was
found to be roughly compatible with the CERES data \cite{Cass95C,Rapp},
too.  On the other hand, the dimuon data of the HELIOS-3
collaboration~\cite{HELIOS} could only be described satisfactorily when
including 'dropping' meson masses~\cite{Cass96H,Li96}.

Though meanwhile there are a couple of hints on 'dropping' meson masses
with baryon density and thus on a partial restoration of chiral
symmetry for very hot and dense nuclear matter, a more systematic study
of medium effects on the differential dilepton spectra -- also with
respect to the transverse momentum $p_T$ and rapidity $y$ -- appears
necessary to optimize the experimental setups.

Our paper is organized as follows: we start with a brief reminder of
the covariant transport approach employed in our analysis in Sect. 2
and present the actual expressions and form factors used for the
evaluation of the differential dilepton multiplicities.  In Sect.~3 we
investigate the question, at which bombarding energies one might find
optimal conditions for high baryon densities at sufficiently large
timescales in Pb~+~Pb collisions.  Section 4 contains the comparison of
our calculated dilepton spectra for p~+~Be and Pb~+~Au reactions with
experimental data at SPS energies. Detailed predictions for dilepton
production form AGS to SPS energies will be presented in Sect.~5 for
p~+~Be and Pb~+~Pb collisions within the 'free' and 'dropping' meson
mass scenario.  Section 6 concentrates on direct photon production in
S~+~Au collisions at 200~GeV/u, while Section 7 concludes our study
with a summary and discussion of open problems.

\section{The covariant transport approach}

In this paper we perform our analysis along the line of the
HSD\footnote{Hadron String Dynamics} approach \cite{Ehehalt} which is
based on a coupled set of covariant transport equations for the
phase-space distributions $f_{h} (x,p)$ of hadron $h$
\cite{Ehehalt,Weber1}, i.e.
\begin{eqnarray}  \label{g24}
\lefteqn{\left\{ \left( \Pi_{\mu}-\Pi_{\nu}\partial_{\mu}^p U_{h}^{\nu}
-M_{h}^*\partial^p_{\mu} U_{h}^{S} \right)\partial_x^{\mu}
+ \left( \Pi_{\nu} \partial^x_{\mu} U^{\nu}_{h}+
M^*_{h} \partial^x_{\mu}U^{S}_{h}\right) \partial^{\mu}_p
\right\} f_{h}(x,p) } \nonumber \\
&& = \sum_{h_2 h_3 h_4\ldots} \int d2 d3 d4 \ldots
 [G^{\dagger}G]_{12\to 34\ldots}
\delta^4(\Pi +\Pi_2-\Pi_3-\Pi_4 \ldots )  \nonumber\\
&& \times \left\{ f_{h_3}(x,p_3)f_{h_4}(x,p_4)\bar{f}_{h}(x,p)
\bar{f}_{h_2}(x,p_2)\right.  \nonumber\\
&& -\left. f_{h}(x,p)f_{h_2}(x,p_2)\bar{f}_{h_3}(x,p_3)
\bar{f}_{h_4}(x,p_4) \right\} \ldots\ \ .
\end{eqnarray}
In Eq.~(\ref{g24}) $U_{h}^{S}(x,p)$ and $U_{h}^{\mu}(x,p)$ denote the
real part of the scalar and vector hadron selfenergies, respectively,
while $[G^+G]_{12\to 34\ldots} \delta^4_{\Gamma} (\Pi
+\Pi_2-\Pi_3-\Pi_4\ldots )$ is the 'transition rate' for the process
$1+2\to 3+4+\ldots$ which is taken to be on-shell in the
semiclassical limit adopted. The hadron quasi-particle properties in
(\ref{g24}) are defined via the mass-shell constraint \cite{Weber1},
\begin{equation}   \label{g25}
\delta (\Pi_{\mu}\Pi^{\mu}-M_{h}^{*2} ) \ \ ,
\end{equation}
with effective masses and momenta given by
\begin{eqnarray} \label{g26}
M_{h}^* (x,p)&=&M_h + U_h^{{S}}(x,p) \nonumber \\
\Pi^{\mu} (x,p)&=&p^{\mu}-U^{\mu}_h (x,p)\ \ ,
\end{eqnarray}
while the phase-space factors
\begin{equation}
\bar{f}_{h} (x,p)=1 \pm f_{{h}} (x,p)
\end{equation}
are responsible for fermion Pauli-blocking or Bose enhancement,
respectively, depending on the type of hadron in the final/initial
channel. The dots in Eq.~(\ref{g24}) stand for further contributions
to the collision term with more than two hadrons in the final/initial
channels. The transport approach (\ref{g24}) is fully specified by
$U_{h}^{S}(x,p)$ and $U_{h}^{\mu}(x,p)$ $(\mu =0,1,2,3)$, which
determine the mean-field propagation of the hadrons, and by the
transition rates $G^\dagger G\,\delta^4 (\ldots )$ in the collision
term, that describes the scattering and hadron production/absorption
rates.

\subsection{Hadron selfenergies}

The scalar and vector mean fields $U_{h}^{S}$ and $U^\mu_{h}$ for
baryons are taken from Ref.~\cite{Ehehalt} and don't have to be
specified here again, since variations in the baryon selfenergies
within the constraints provided by experimental data were found to have
no sizeable effect on the dilepton differential spectra.
In the present approach we propagate explicitly -- apart from the
baryons (cf.~\cite{Ehehalt}) -- pions, kaons, $\eta$'s, $\eta^\prime$'s
the $1^-$ vector mesons $\rho, \omega, \Phi$ and $K^*$'s as well as the
axial vector meson $a_1$. We
assume that the pions as Goldstone bosons do not change their
properties in the medium; we also discard selfenergies for the
$\eta$ and $\eta^\prime$-mesons in the present calculation, since we did not 
find any
appreciable selfenergy effects in comparison to the experimental
spectra available in the energy regime of interest here.

The kaon selfenergies are described as in Ref.~\cite{Ehehalt}
following Kaplan and Nelson~\cite{Kaplan}, i.e. the kaon masses ($K^+,
K^0$) are assumed not to be changed in the medium due to an approximate
cancellation of attractive scalar and repulsive vector interactions,
whereas the antikaons drop in mass according to
\begin{equation}
m^*_{\bar{K}} = m^0_{\bar{K}}\ \left(1 - \lambda_{\bar{K}}
{\rho_B\over \rho_0}\right)  \geq (m_u + m_s) \approx 0.16 \ {\rm GeV}
\label{kaons}
\end{equation}
with $\lambda_{\bar{K}} \approx$ 0.16 as in Ref.~\cite{Ehehalt}.  In
(\ref{kaons}) $m^0_{\bar{K}}$ and $\rho_0 \approx 0.16 \ {\rm fm}^{-3}$
denote the vacuum antikaon mass and the saturation density of nuclear
matter, respectively. Due to a lack of further knowledge about the
$K^*$ vector mesons we assume the same scaling with baryon density as
for the kaons\footnote{This working hypothesis needs to be controlled
in future.}.

The in-medium properties of the vector mesons $\rho, \omega$ and $\Phi$
are modelled according to the QCD sum rule analysis by Hatsuda and Lee
\cite{hatsuda} as in our previous studies, i.e.
\begin{equation}
m^*_{V} = m^0_{V}\  \left(1 - \lambda_V \frac{\rho_B}{\rho_0}\right)
\geq (m_u + m_d) \approx 0.016 \ {\rm GeV}
\label{vector}
\end{equation}
with $\lambda_V \approx$ 0.18 for $\rho$ and $\omega$, while the $\Phi$
meson is expected to scale with $\lambda_V \approx$ 0.025. In line with
(\ref{vector}) the in-medium $\Phi$ mass is limited to twice the strange
quark mass, i.e.  $m_\Phi \geq 2 m_s \approx 0.3$~GeV.

In addition to our previous works \cite{Ehehalt,Cass95C,Cass96H,BCM96an}
we now also include the axial vector meson $a_1$, i.e. the chiral
partner of the $\rho$ meson. Since the scaling of the $a_1$ mass with
baryon density is quite a matter of debate and we only know that
$m^*_{a_1} = m^*_\rho$ in the chiral limit, we assume the $a_1$ to
scale in mass in the same way as the $\rho$ meson according to
Eq.~(\ref{vector}) in order not to introduce any new parameter.

In the following, the 'free' meson mass scenario will employ the limit
$\lambda_{\bar{K}} = \lambda_V = \lambda_{a_1}$ = 0, whereas the
'dropping' mass scenario is described by Eqs.~(\ref{kaons}) and
(\ref{vector}), respectively.

\subsection{Dilepton channels}

In this analysis we calculate dilepton production by taking into account
the contributions from the Dalitz-decays $\eta \to \gamma e^+e^-$,
$\omega \to \pi^0 e^+e^-$, $\eta^\prime \to \gamma e^+e^-$,
$a_1 \to \pi e^+e^-$and the direct dilepton decays of the vector mesons
$\rho, \omega, \Phi$ where the $\rho, \Phi$ and $a_1$ mesons may as well
be produced in $\pi \pi$, $K \bar{K}$ and $\pi \rho$ collisions, respectively.

In case of a perturbative treatment\footnote{The perturbative treatment is
used to test the dynamical scheme described below in case of the bare 
$\rho$-mass
scenario, only.} of the channel $\pi^+ \pi^- \rightarrow \rho^0
\rightarrow e^+ e^-$
the cross section is parametrized as in Refs.~\cite{Wolf1,Wolf3,GaleK87} by
\begin{equation}
\sigma_{\pi^+\pi^-\to e^+e^-}(M) = {4\pi\over 3} \
 {\alpha^2\over M^2} \ \sqrt{1- {4 m_\pi^2\over M^2}}\ |F_\pi(M)|^2,
\label{piann}
\end{equation}
where the 'free' form factor of the pion is approximated by
\begin{equation}
|F_\pi(M)|^2 = {m_\rho^4\over (M^2-m_\rho {}^2)^2+m_\rho^2 \Gamma_\rho^2}.
\end{equation}
In Eq.~(\ref{piann}) $M$ is the dilepton invariant mass, $\alpha$ is
the fine structure constant, and
$$ m_\rho = 775\ {\rm MeV},  \ \ \Gamma_\rho = 118\ {\rm MeV}. $$

The cross section for $K^+K^-$ annihilation is parametrized as \cite{liko}
\begin{equation}
\sigma_{K^+K^-\to e^+e^-}(M) = {4\pi\over 3} \
\left({\alpha\over M} \right)^2 \ \sqrt{1-{4 m_K^2\over M^2}} \ |F_K(M)|^2,
\label{skaon}
\end{equation}
where the 'free' form factor of the kaon is approximated by
\begin{equation}
|F_K(M)|^2 = {1\over 9} {m_\Phi^4\over (M^2-m_\Phi^2)^2+m_\Phi^2
\Gamma_\Phi^2}
\label{fkaon}
\end{equation}
with
$$m_\Phi = 1020\ {\rm MeV},  \ \ \Gamma_\Phi = 4.43\ {\rm MeV}.$$

The cross section for dilepton production in $\pi^+\rho^- \to \Phi \to
e^+e^-, \ \pi^-\rho^+\to \Phi \to e^+e^-$ scattering can be represented as
\begin{eqnarray}
\sigma_{\pi\rho\to \Phi\to e^+e^-}(M) = {1\over 3} \ {B_{\rho\pi}\over
B_{K^+K^-}} \ \sigma_{K^+K^-\to \Phi\to e^+e^-}
\label{spirho}\end{eqnarray}
with $B_{\rho\pi}=0.13, \ \ B_{K^+K^-}=0.49$.

The $\eta$ Dalitz decay is given by  \cite{Landsberg}:
\begin{eqnarray}
{d\Gamma_{\eta \to \gamma e^+e^-}\over dM}
= {4\alpha\over 3\pi} \ {\Gamma_{\eta \to 2\gamma}\over M}
\left(1 - {4 m_e^2\over M^2} \right)^{1/2} \left( 1 + 2 {m_e^2\over M^2}
\right)   \nonumber \\
\times \left( 1 - {M^2\over m_\eta^2}\right)^3
\ |F_{\eta \to \gamma e^+e^-}(M)|^2,
\label{gameta}
\end{eqnarray}
where the form factor is parametrized in the pole approximation as
\begin{eqnarray}
F_\eta (M) = \left(1-{M^2\over \Lambda_\eta^2}\right)^{-1},
\label{feta}\end{eqnarray}
with the cut-off parameter $\Lambda_\eta \simeq 0.72$~GeV.

Similarly, the $\omega$ Dalitz-decay is \cite{Landsberg}:
\begin{eqnarray}
&& {d\Gamma_{\omega \to \pi^0 e^+e^-}\over dM}
= {2\alpha\over 3\pi} \ {\Gamma_{\omega \to \pi^0 \gamma}\over M}
\left(1 - {4 m_e^2\over M^2} \right)^{1/2} \left( 1 + 2 {m_e^2\over M^2}
\right)   \nonumber \\
&&\times \left[ \left(1+ {M^2\over m_\omega^2 - m_\pi^2}\right)^2 -
{4 m_\omega^2 M^2\over (m_\omega^2-m_\pi^2)^2} \right]^{3/2} \
|F_{\omega \to \pi^0 e^+e^-}(M)|^2,
\label{gamomeg}
\end{eqnarray}
where the form factor squared is parametrized as
\begin{equation}
|F_{\omega \to \pi^0 e^+e^-}(M)|^2 = {\Lambda_\omega^4\over
(\Lambda_\omega^2- M^2)^2 + \Lambda_\omega^2 \Gamma_\omega^2}
\label{fomegD}
\end{equation}
with
$$ \Lambda_\omega= 0.65\ {\rm GeV}, \ \ \Gamma_\omega = 75\ {\rm MeV}.$$

For $\eta^\prime \to \gamma e^+e^-$ we use a similar expression as
Eq.~(\ref{gameta}). However, in this case the pole approximation
(\ref{feta}) is no longer valid since the vector meson pole occurs in the
physical region of the dilepton spectrum ($M < m_{\eta^\prime}$).
Instead we use a form factor of the form (\ref{fomegD})
with
$$ \Lambda_{\eta^\prime}= 0.75\ {\rm GeV}, \ \ \Gamma_{\eta^\prime}
= 0.14\ {\rm GeV},$$
which reproduces the experimental data from \cite{Landsberg} reasonably well.

The direct decays of the vector mesons $\omega, \Phi$
to $e^+e^-$ are taken as
\begin{equation}
\frac{d\sigma}{dM^2}(M) = {1\over \pi} {m_V \Gamma_V\over (M^2-m_V^2)^2
+ m_V^2 \Gamma_V^2} {\Gamma_{V \to e^+e^-}\over \Gamma_V}
\end{equation}
with $\Gamma_{\Phi \to e^+e^-}/\Gamma_\Phi= 2.5\times 10^{-4}$ and
$\Gamma_{\omega \to e^+e^-}/\Gamma_\omega = 7.1\times 10^{-5}$.

The dilepton channels $\eta \rightarrow e^+ e^-, \ \omega \rightarrow
\pi^0 e^+ e^-, \ \omega \rightarrow e^+ e^-, \ \Phi \rightarrow e^+ e^-, \
\pi \rho \rightarrow \Phi \rightarrow e^+ e^-$ and $K^+ K^- \rightarrow
\Phi \rightarrow e^+ e^-$ are treated perturbatively and are computed
at the end of the transport calculation
due to the 'long' lifetime of the mesons
$\eta, \eta^\prime, \omega, \Phi$. The decays of the short-lived $\rho$ and
$a_1$ mesons, however, have to be treated explicitly since these mesons
change their properties rapidly (in case of the dropping mass scenario) and
decay during the expansion phase of the system.

The mesons ($\rho, a_1$) stemming from a string decay with invariant
mass $m^*_V$ at baryon density $\rho_B$ according to Eq.~(\ref{vector})
are selected by Monte Carlo according to the Breit-Wigner distribution:
\begin{equation}
f(M) =  N_V \ {2\over \pi} \ {M m_V^* \Gamma_{V}^{*}
\over (M^2-m_{V}^{*2})^2 + m_V^{*2} \Gamma_{V}^{*2}},
\end{equation}
where $N_V$ guarantees normalization to unity, i.e. $\int f(M) dM =1$.
The width $\Gamma_V^*(M)$ is determined from
\begin{equation}
\Gamma_V^*(M) = \Gamma_0 \left({m_{V}^*\over M}\right)^2  \
{q\over q_V},
\end{equation}
where $\Gamma_0$ is the full width at the mean resonance energy, $q$
and $q_V$ are the pion three-momenta in the restframe of the resonance
with mass $M$ and  $m_V^*$, respectively.

Apart from string decay the mesons ($\rho, a_1$) are abundantly also
created from $\pi \pi$ or $\pi \rho$ collisions, respectively.  For
the $a_1$ formation cross section in the reaction $\pi^+ \rho^-\to a_1,
\pi^- \rho^+ \to a_1 $ or for the $\rho^0$ cross section the reaction
$\pi^+ \pi^- \to \rho^0$ we use the Breit-Wigner form
\cite{Xiong92,Ko81}:
\begin{eqnarray}
\sigma_{V}(s) = {2 J_V +1\over 2S +1} \
{4\pi\over k^2} \ { m_V^{*2} \Gamma_V^{*2}
\over (s-m_{V}^{*2})^2 + m_V^{*2} \Gamma_V^{*2} },
\label{a1prod}\end{eqnarray}
where $k$ is the pion momentum in the center-of-mass of the produced
meson $V$ = ($\rho, a_1$), $s$ is the invariant energy squared while
$J_V$ stands for the spin of the produced meson and $S$ for the spin of
the collision partner of the pion, respectively. In the time reversed
processes the vector mesons of actual mass $M$ may decay in each
timestep according to the probability
\begin{equation}
P = {\rm exp}(-\Gamma_V^*(M)/\gamma \Delta t)
\end{equation}
where $\Delta t$ is the actual timestep size and $\gamma$ the Lorentz
factor of the resonance with respect to the calculational frame.

The $\rho^0$ decay to $e^+ e^-$ with invariant mass $M$ is calculated
by integrating the equation (using the mass bin $\Delta M$)
\begin{equation}
{d N_{e^+ e^-}^{\rho}\over \Delta M dt} = {1\over \hbar}
\sum_{{\rm events} \ i=1}^{N_{\rho^0}(M,t)} \Gamma_{\rho^0 \to e^+ e^-} (M)
{1\over \Delta M}
\label{Nrho}
\end{equation}
in time with
\begin{equation}
\Gamma_{\rho^0 \to e^+ e^-}(M) = 8.8 \times 10^{-6} \ M,
\label{gamrn}
\end{equation}
where $N_{\rho^0} (M,t)$ is the number of $\rho^0$ mesons of mass $M$
at time $t$ in the calculation. The factor $8.8 \times 10^{-6}$ stems
from the measured width of the $\rho^0$ to $e^+ e^-$.  It can be shown
\cite{liko} that the method described above leads to the same result as
Eq.~(\ref{piann}) if the $\rho^0$ meson does not change its properties
in time. We note that by treating explicitely the $\rho$ formation by $\pi \pi$
collisions the perturbative channel (\ref{piann})
has been switched off to avoid double counting.

The $a_1$ decay to $\pi e^+ e^-$ with invariant mass $M$ is calculated in
analogy by integrating the equation
\begin{equation}
{d N_{e^+ e^-}^{a_1}\over dM dt} = {1\over \hbar}
\sum_{{\rm events}\ i=1}^{N_{a_1} (t)}
{d\Gamma_{a_1 \to \pi e^+ e^-} \over dM}(m_{a_1}^*,M)
\label{Na1}
\end{equation}
in time with
\begin{eqnarray}
 {d\Gamma_{a_1 \to \pi e^+ e^-}\over dM}(m_{a_1}^*,M) =
{2 \alpha \over 3 \pi} \ {\Gamma_{a_1\to \pi\gamma}\over M} \
\left(1 - {4 m_e^2\over M^2} \right)^{1/2} \left( 1 + 2 {m_e^2\over M^2}
\right)   \nonumber \\
 \times \left[\left(1 + {M^2\over m_{a_1}^{*2} - m_{\pi}^2}\right)^2
- {4 m_{a_1}^{*2} M^2\over (m_{a_1}^{*2} - m_{\pi}^2)^2} \right]^{3/2}
\label{gama_a1}
\end{eqnarray}
with $\Gamma_{a_1\to \pi\gamma} = 6.4\times 10^{-4}$~GeV, while
$N_{a_1} (t)$ is the number of $a_1$ mesons at time $t$ in the
calculation.

We note that we discard baryon-baryon ($BB$), meson-baryon ($mB$) and
meson-meson ($mm$) bremsstrahlung channels as well as the Dalitz decays
of the baryon resonances since their contribution was found to be small
in Refs.~\cite{Cass95C,Cass96H}.

Before going over to the actual calculations for $e^+ e^-$ spectra, we
start with some more general analysis of central heavy-ion collisions from
1 -- 200~GeV/u.

\section{Optimizing for high baryon density}

\begin{figure}[h]
\vspace*{-2cm}
{\psfig{figure=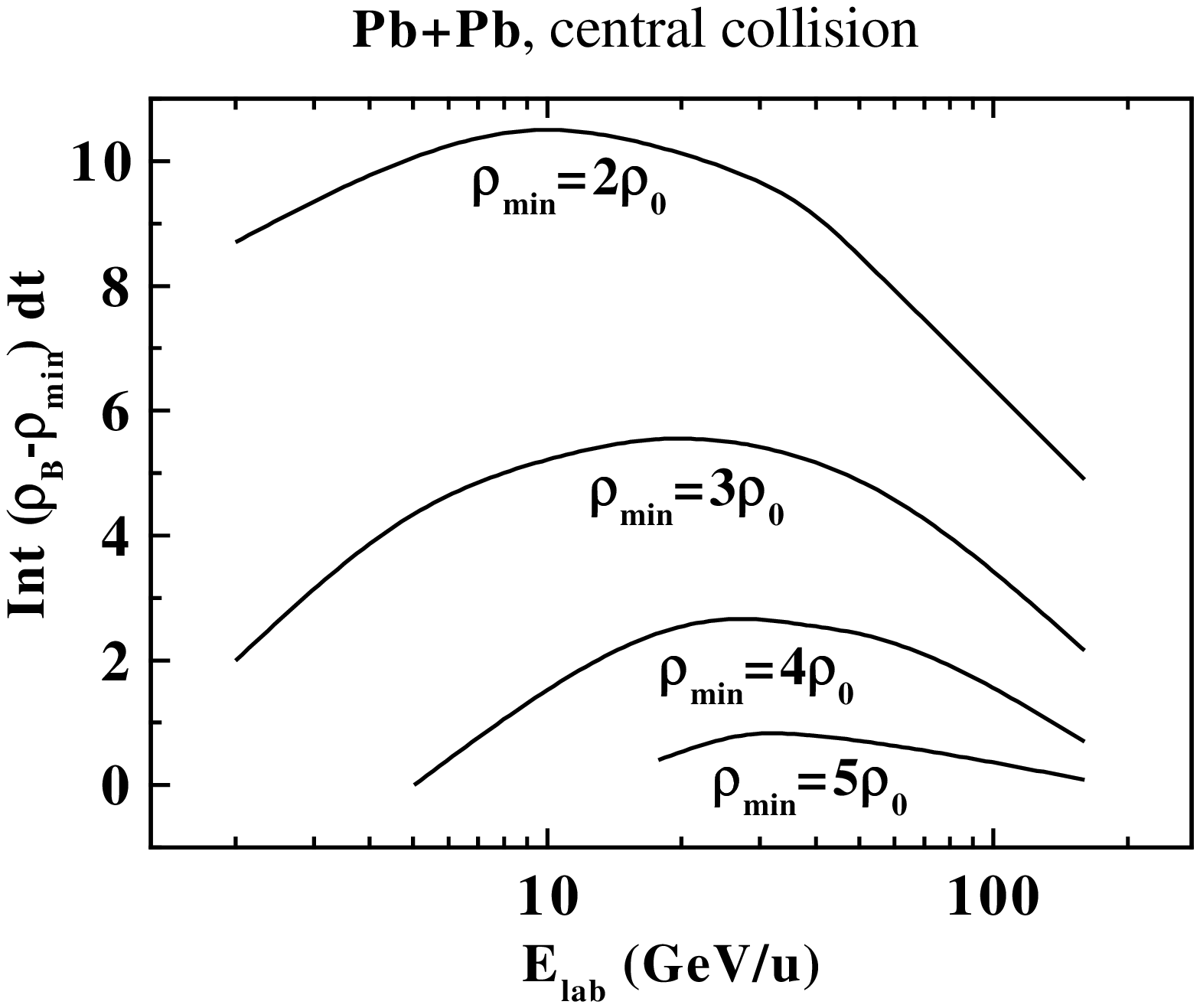,width=15cm,height=22cm}}
\vspace*{-2cm}
\caption{The quantity F (Eq.~(\protect\ref{e1})) for central Pb~+~Pb
collisions as a function of the bombarding energy per nucleon for
4 different cuts in $\rho_{min}$.}
\label{Fig1}
\end{figure}

In order to probe the restoration of chiral symmetry at high baryon
density in nucleus-nucleus collisions, one has to perform experiments
with heavy nuclei (e.g. Pb~+~Pb) and optimize the beam energy to
achieve a large volume of high baryon density for a sufficiently long
time. In this respect central collisions of Pb~+~Pb have been
investigated within the transport approach specified above and the
'stopped' baryon density $\rho_B^s(t)$ -- including only baryons with
rapidity $|y| \leq 0.7$ in the cms -- has been computed in a central
cylinder of the volume $V = \pi R^2 \Delta_z/\gamma_{cm}$ with
$\Delta_z = R=4$~fm, while $\gamma_{cm}$ is the Lorentz factor in the
nucleus-nucleus center-of-mass system.  Since we are interested in high
baryon densities above some value $\rho_{min}$ for long times, we
consider the quantity
\begin{eqnarray}
F = \int dt \ (\rho_B^s(t) - \rho_{min}) \ \Theta(\rho_B^s(t) - \rho_{min}),
\label{e1}
\end{eqnarray}
which should serve as a useful guide in the optimization problem. The
quantity F (Eq.~(\ref{e1})) is displayed in Fig.~\ref{Fig1} for central
collisions of Pb~+~Pb from 1 -- 200~GeV/u for different values of
$\rho_{min}$ from $2\rho_0-5\rho_0 \ (\rho_0 \approx 0.168\ {\rm
fm}^{-3})$.  Accordingly, optimal bombarding energies for baryon
densities above $4\rho_0$ should be around 20 -- 30~GeV/u in order to
explore the properties of an intermediate phase, where the chiral
symmetry might approximately be restored and the hadron masses (except
for the Goldstone bosons) might be close to their current quark masses
$m_q + m_{\bar{q}}$. However, also lower bombarding energies
(2 -- 10~GeV/u) are seen to qualify for studies of a partial restoration
of chiral symmetry since sizeable space-time volumes with baryon
densities above 2 -- 3~$\rho_0$ can be achieved.

\begin{figure}[h]
\vspace*{-2cm}
{\psfig{figure=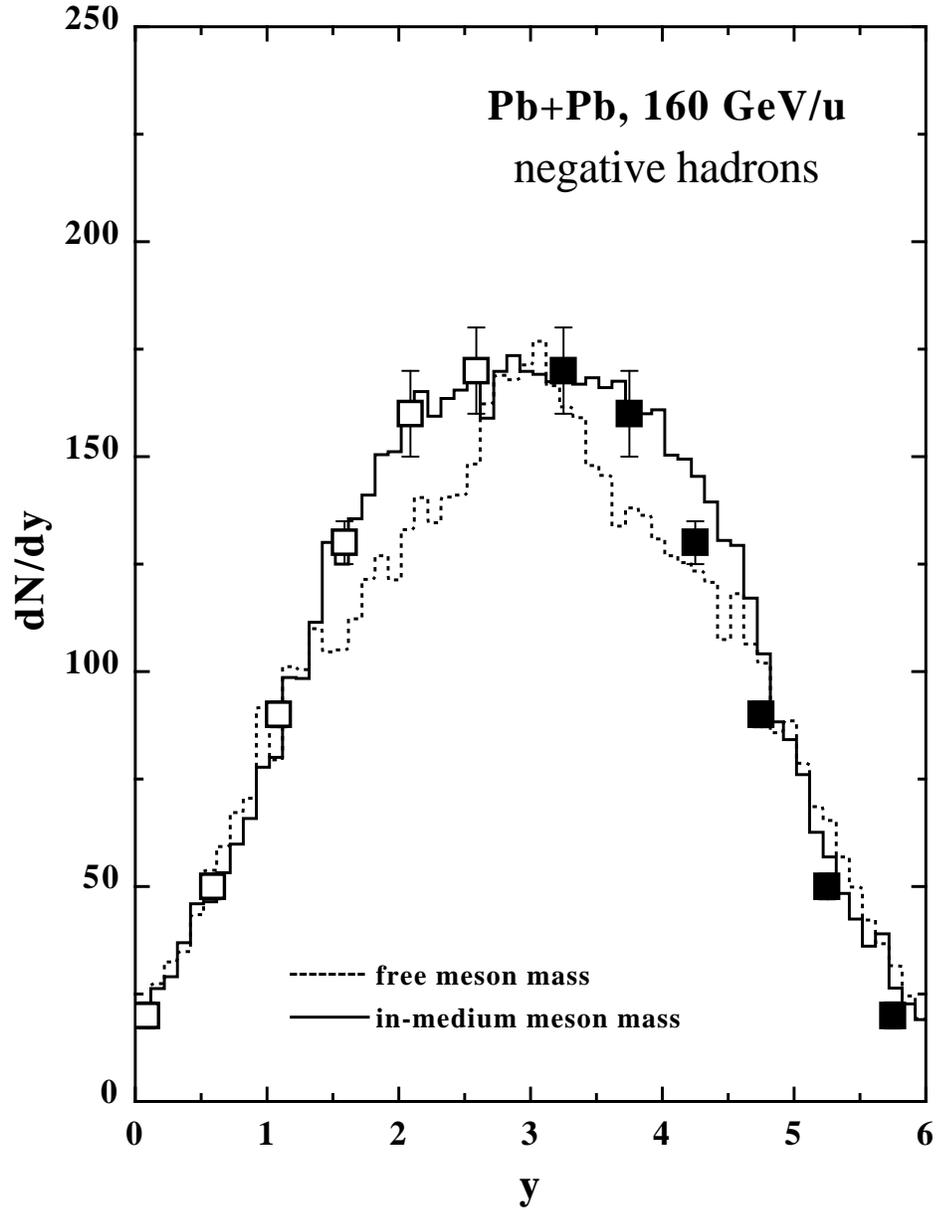,width=15cm,height=22cm}}
\vspace*{-2cm}
\caption{The preliminary rapidity distribution of negative hadrons from
NA49~\protect\cite{NA49} (full squares) in comparison to the HSD
results. The solid line corresponds to a calculation (at $b=2$~fm)
including the 'dropping' meson masses from Eqs.~(\protect\ref{kaons}),
(\protect\ref{vector}), whereas the dotted line results from a
calculation with bare meson masses. The open squares are obtained by
reflecting the full squares at midrapidity.}
\label{Fig2}
\end{figure}

Whereas the HSD approach has been shown in Ref.~\cite{Ehehalt} to
reasonably reproduce hadronic spectra and rapidity distributions for
heavy-ion collisions at SIS and AGS energies, a definite proof for
heavy systems at SPS energies is still lacking though quantitative
predictions for the baryon rapidity distribution in Pb~+~Pb collisions
at 158~GeV/u have been given in Ref.~\cite{Ehehalt}. In order to
demonstrate, that the analysis presented in Fig.~\ref{Fig1} is
meaningful also at higher bombarding energy, we compare in
Fig.~\ref{Fig2} the preliminary rapidity distribution of negative
hadrons (essentially $\pi^-$, $K^-$ and $\bar{p}$) from
NA49~\cite{NA49} for central Pb~+~Pb collisions at 160~GeV/u with the
HSD results.  Here the solid line corresponds to a calculation (at $b =
2$~fm) including the 'dropping' meson masses from Eqs.~(\ref{kaons}),
(\ref{vector}), whereas the dotted line results from a calculation with
bare meson masses. The broadening of the rapidity distribution around
midrapidity ($y\approx 3$) in the 'dropping' mass scenario is due to
pions from $\rho$ and $\omega$ decays, which are produced with a wider
distribution in rapidity in this case. The quantitative agreement with
the data, however, indicates that both scenarios -- i.e. 'free' and
'dropping' meson masses as described above -- are compatible with the
present preliminary data. We note that the open squares in
Fig.~\ref{Fig2} have been obtained by reflecting the full squares (from
NA49) at midrapidity.  The fact, that the calculated spectrum is not
fully symmetric with respect to midrapidity, is caused by the finite
statistics of the computation, which was performed with 150 parallel
runs in this case.

\begin{figure}[h]
\vspace*{-2cm}
{\psfig{figure=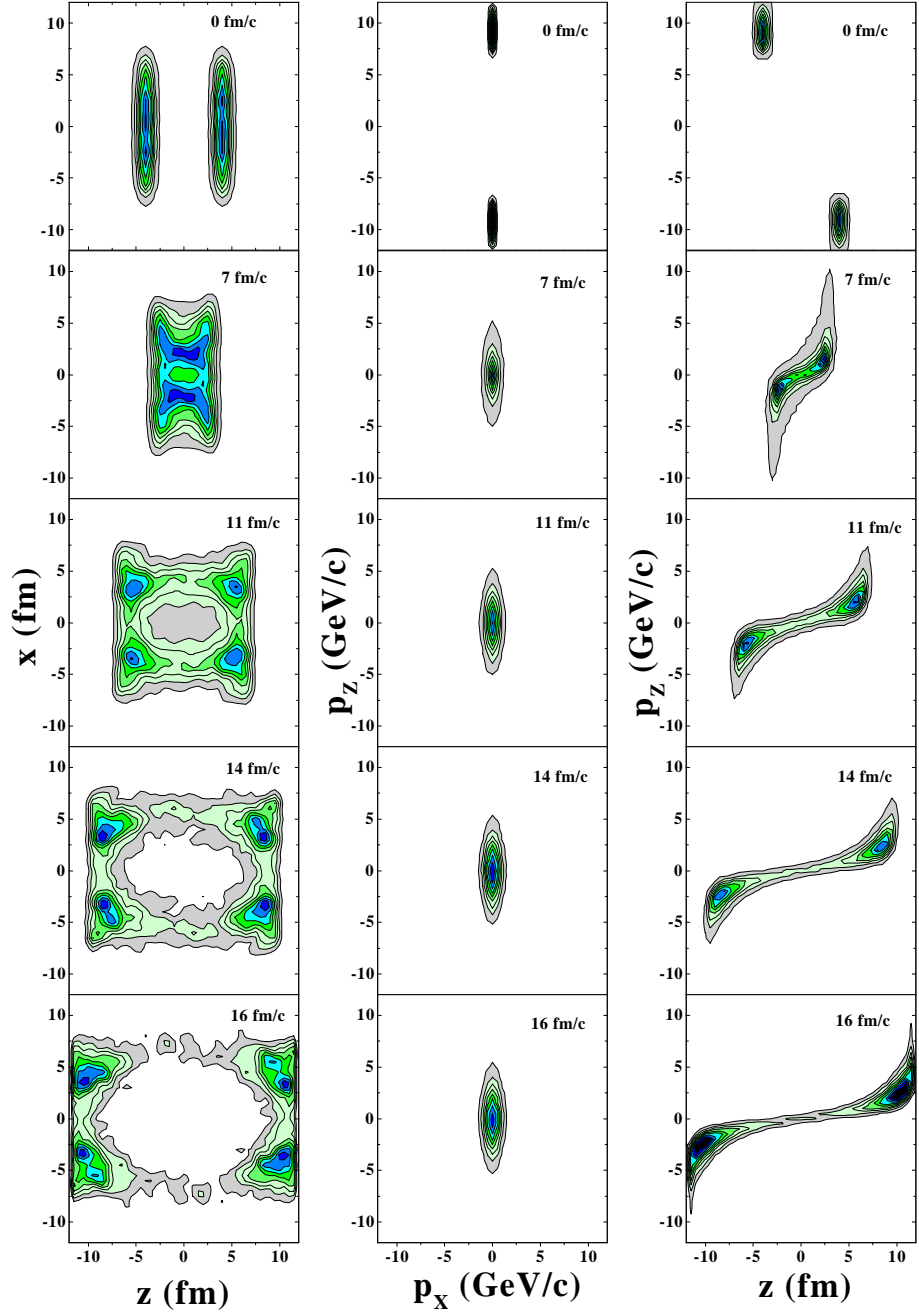,width=15cm,height=22cm}}
\vspace*{-2cm}
\caption{Baryon density distribution (left column), momentum space
(middle column) and phase-space distribution (right column) for
a 160~GeV/u Pb~+~Pb collision at $b=0$~fm for various times in fm/c.}
\label{Fig3}
\end{figure}

Since the system Pb~+~Pb at 160~GeV/u is explored experimentally at the
SPS in great detail, it is advantages to have a look at the space-time
evolution of the baryon one-body density for central collisions. In
this respect we show in Fig.~\ref{Fig3} the space-time evolution of
baryons for this system at $b = 0$~fm:  (l.h.s.)  countor plot of the
baryon density distribution in coordinate space $\rho_B(x,y=0,z;t)$,
(middle colum) contour
plot of the baryon momentum distribution $\rho_B(p_x,p_y=0,p_z;t)$,
(r.h.s.) the phase-space distribution
\begin{eqnarray}
f(z, p_z; t) = (2\pi)^{-2} \ \sum\limits_b \int dr_\perp dp_\perp \
f_b (r_\perp, z, p_\perp, p_z; t),
\label{fzpzt}\end{eqnarray}
where $\sum\limits_b$ denotes a sum over all baryon species. In the
time evolution of the density distribution $\rho_B(x,y=0,z;t)$ we
explicitly mention the short phase of high baryon density from about
5 -- 8~fm/c as well as the sizeable fraction of 'spectators' from the
nuclear corona.  The time evolution in momentum space (middle colum)
shows that the system reaches its final distribution within a few fm/c,
however, is far from the kinetic equilibrium in the baryon degrees of
freedom, which would be reflected by a spherical distribution here.
It clearly indicates a dominant longitudinal expansion of the system,
which is much more pronounced than at AGS energies (cf. Fig.~21 in
\cite{Ehehalt}).

\section{Comparison with experimental dilepton data at SPS energies}

The covariant transport approach HSD has been applied already to the
analysis of dilepton spectra for p~+~Be and Ca~+~Ca collisions at
BEVALAC/SIS energies \cite{brat96}, for p~+~Be, p~+~Au, and S~+~Au
collisions at SPS energies~\cite{Cass95C} and to dimuon spectra for
p~+~W and S~+~W at 200~GeV/u~\cite{HELIOS} in Ref.~\cite{Cass96H}.
Whereas the dilepton spectra for p~+~A reactions at high energy could
be well described by the mesonic decays in line with the experimental
analysis~\cite{CERES}, the S~+~A data could only be satisfactorily
reproduced within the 'dropping' mass scenario (cf. also
\cite{Li,Li96}). Here, we include additional reaction channels (e.g.
the $\eta^\prime, a_1$ degrees of freedom) and present a systematic
study with respect to differential spectra in the transverse momentum
of the lepton pair, the rapidity distribution as well as the dilepton
yield as a function of the charged particle multiplicity at
midrapidity. Before doing so, we present the actual status of our
calculations in comparison with presently available data on dilepton
production at SPS energies.

\subsection{Differential dilepton spectra for p~+~Be and Pb~+~Au
at SPS energies}

\begin{figure}[h]
\vspace*{-2cm}
{\psfig{figure=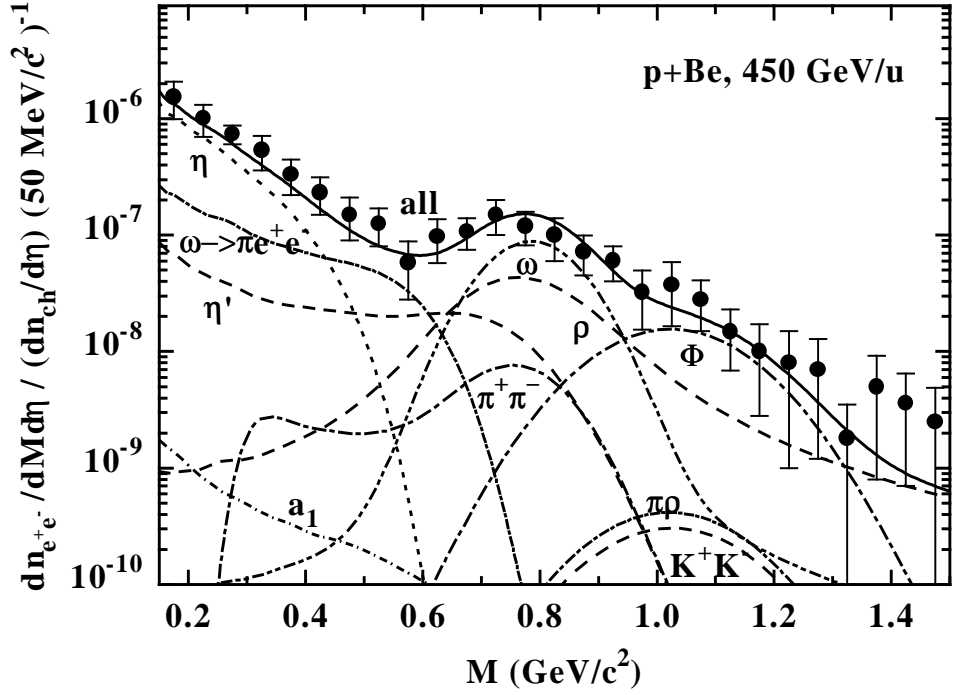,width=15cm,height=22cm}}
\vspace*{-2cm}
\caption{The calculated dilepton spectra (full solid line) for p~+~Be
at 450~GeV/u in comparison with the data from
Ref.~\protect\cite{CERES}.  The thin lines indicate the individual
contributions from the different production channels including the
CERES-acceptance and mass resolution; i.e. starting at low $M$: $\eta
\to \gamma e^+ e^-$ (dashed line), $\omega \to \pi^0 e^+ e^-$
(dot-dashed line), $\eta^\prime \to \gamma e^+ e^-$ (long dashed line),
$a_1 \to \pi e^+e^-$ (dot-dashed line); for $M \approx $ 0.8 GeV:
$\omega \to e^+e^-$ (dot-dot-dashed line), $\rho^0 \to e^+e^-$ (dashed
line), $\pi^+ \pi^- \to e^+e^-$ (dot-dashed line); for $M \approx $ 1
GeV: $\Phi \to e^+e^-$ (dot-dashed line), $\pi \rho \to e^+e^-$
(dot-long dashed line), $K \bar{K} \to e^+e^-$ (dashed line).}
\label{Fig4}
\end{figure}

As an example for dilepton spectra at SPS energies Fig.~\ref{Fig4}
shows the spectral decomposition as a function of the $e^+e^-$
invariant mass M for p~+~Be at 450~GeV/c in comparison to the data of
the CERES collaboration~\cite{CERES}.  The $e^+, e^-$ acceptance cuts
in pseudo-rapidity ($2.1 \le \eta \le 2.65$), a cut of the transverse
$e^+$ and $e^-$ momenta for $p_T \ge 0.05$~GeV/c as well as a cut on
the opening angle of the $e^+e^-$ pair ($\Theta \ge 35$~mrad) are taken
into account. Furthermore, the experimental mass resolution has been
included in evaluating the final mass spectrum, which is normalized by
the number of charged particles $dn_{ch}/d\eta$ in the pseudorapidity
bin 2.1 $\leq \eta \leq$ 3.1.  As can be extracted from
Fig.~\ref{Fig4}, the spectrum for p~+~Be can be fully accounted for by
the electromagnetic decays of the $\eta, \eta^\prime, a_1$ and vector
mesons $ \rho^0, \omega$ and $\Phi$; contributions from meson-meson
channels ($\pi^+\pi^-, K^+K^-, \pi \rho$) are of minor importance here.
We note that in our present analysis  the $a_1$ and $\eta^\prime$
Dalitz decays have been taken into account in addition to
Ref.~\cite{Cass95C}.  The $\eta^\prime$ contribution is more sensitive
to the electromagnetic form factor then $\eta$ or $\omega$ Dalitz
decays because the vector meson pole shows up in the region $M \le
m_{\eta^\prime}$.  However, the contribution of the $\eta^\prime$
Dalitz decays is not essential compared to the other channels; the
$a_1$ contribution is practically negligible for p~+~Be.

\begin{figure}[h]
\vspace*{-3cm}
{\psfig{figure=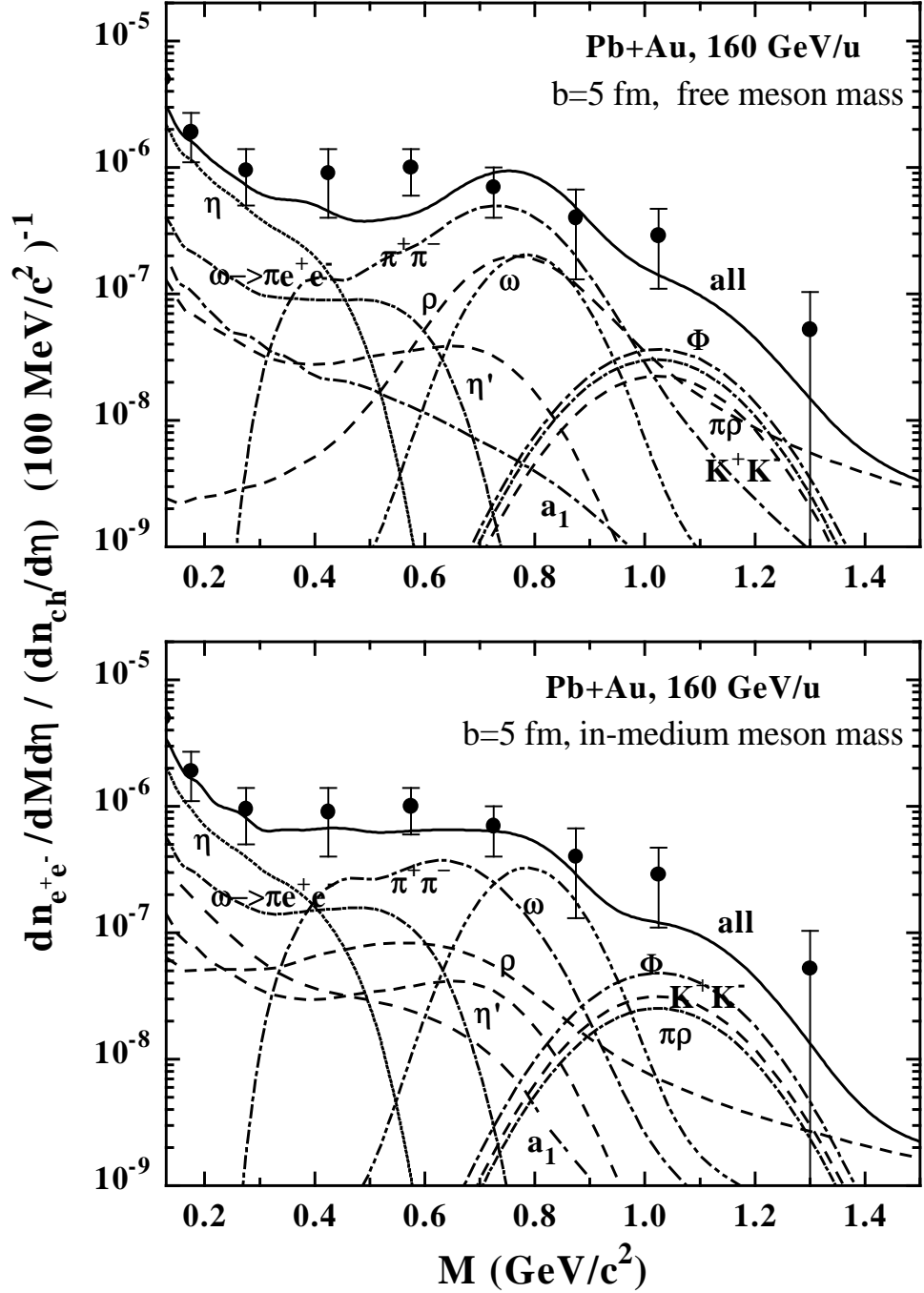,width=15cm,height=22cm}}
\vspace*{-.5cm}
\caption{Dilepton invariant mass spectra for semicentral collisions of
Pb~+~Au at 160~GeV/u (full solid lines) in comparison to the
preliminary data of the CERES collaboration \protect\cite{Ullrich}. The
upper part shows the results of a calculation with bare meson masses
whereas the lower part includes the 'dropping' meson masses
(Eqs.~(\protect\ref{kaons}),(\protect\ref{vector})).
The assignment of
the individual contributions is the same as in Fig. 4.}
\label{Fig5}
\end{figure}

The situation changes quite dramatically when going over to
nucleus-nucleus collisions.  In Fig.~\ref{Fig5} we compare the results
of our calculation for the differential dilepton spectra for Pb~+~Au at
160~GeV/u at $b=5$~fm with the preliminary experimental
data~\cite{Ullrich}.  Contrary to p~+~Be reactions, a  cut of the
transverse $e^+, e^-$ momenta $p_T \ge 0.175$~GeV/c has been taken in
line with the experimental acceptance cut.  For Pb~+~Au at 160~GeV/u (and
semicentral collisions) the dominant yield for invariant masses 0.3~GeV
$\leq M$ $\leq$ 0.7~GeV stems from $\pi^+\pi^-$ annihilation (cf.
Fig.~\ref{Fig5}). Also in the $\Phi$ mass regime about 1~GeV there is a
large contribution from $K^+K^-$ and $\pi \rho$ annihilation to
dileptons for both scenarios: with bare meson masses (upper part of
Fig.~\ref{Fig5}) and with in-medium meson masses (lower part of
Fig.~\ref{Fig5}). Whereas most of the processes (Dalitz and direct
decays) occur in the vacuum at zero baryon density, the $\pi \pi \to
\rho^0 \to e^+e^-$ and direct $\rho^0$ (from baryon-baryon  and
meson-baryon collisions) decay still occur at finite baryon density
such that a dropping $\rho$ mass also leads to a shift of the
respective contribution to lower invariant masses $M$. In
Fig.~\ref{Fig5} both scenarios are compared to the preliminary data of
the CERES collaboration~\cite{Ullrich}; due to the present statistics,
however, there is no unique conclusion since the calculation with bare
meson masses (upper part) also describes the data except for one point
at 0.6~GeV (cf. also Ref.~\cite{Frankfurt}). On the other hand,
the present preliminary data match well
with the calculation including the in-medium meson masses. We note that
the comparison with the data in Fig.~\ref{Fig5} has been performed for
$b = 5$~fm, because for this impact parameter the
charged particle multiplicity $dn_{ch}/d\eta \approx$ 260 as for the
experimental normalization (see below).

\subsection{Dilepton yield versus charged particle multiplicity}

\begin{figure}[h]
\phantom{a}
\vspace*{-5cm}
{\psfig{figure=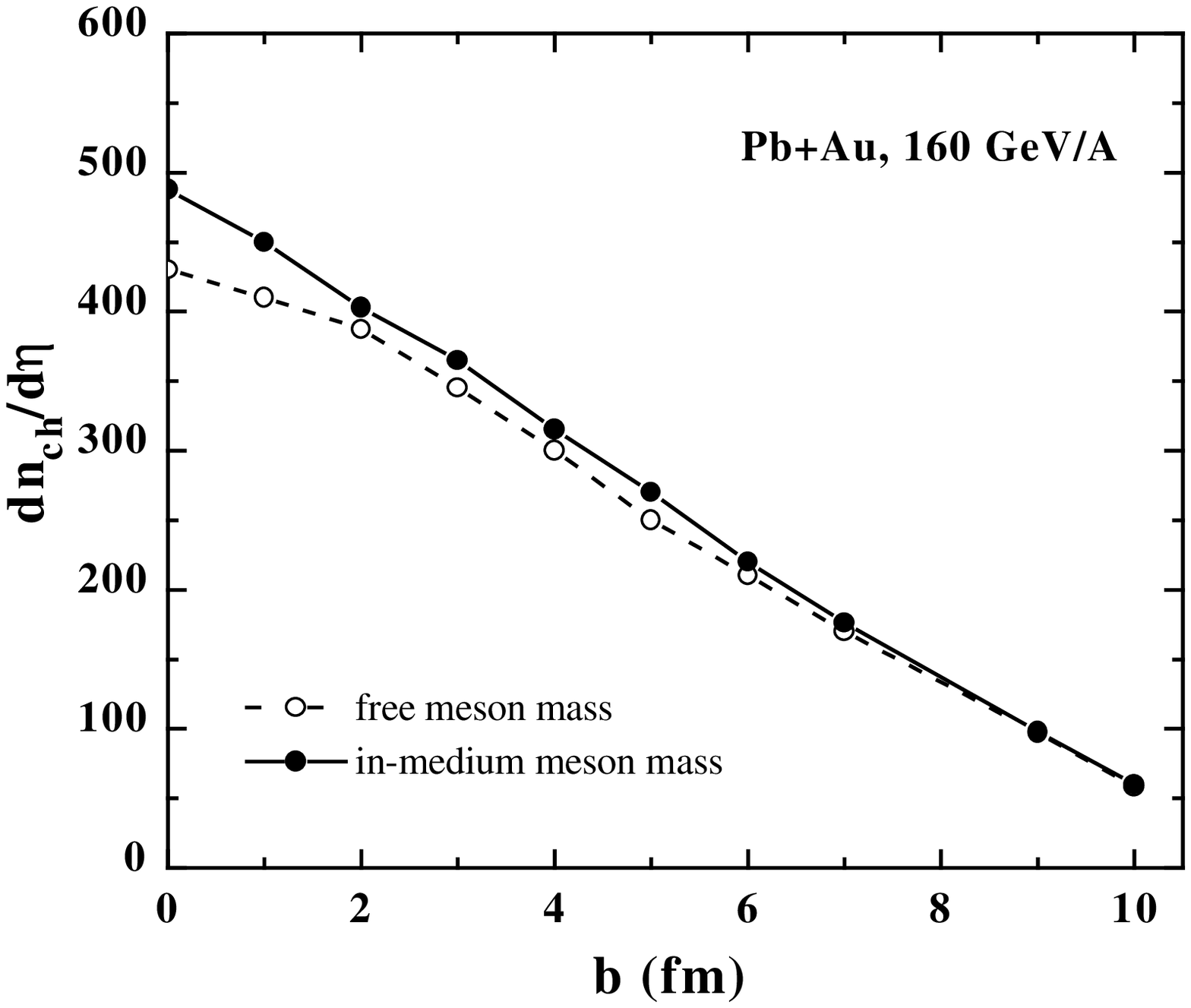,width=15cm,height=18cm}}
\vspace*{-5.3cm}
\caption{The charged particle multiplicity $dn_{ch}/d\eta$
for Pb~+~Au at 160~GeV/u as a function of the impact parameter $b$
for the free meson mass scheme (dashed line) and the dropping meson mass
scenario (solid line).}
\label{Fig6}

\vspace*{-3.5cm}
{\psfig{figure=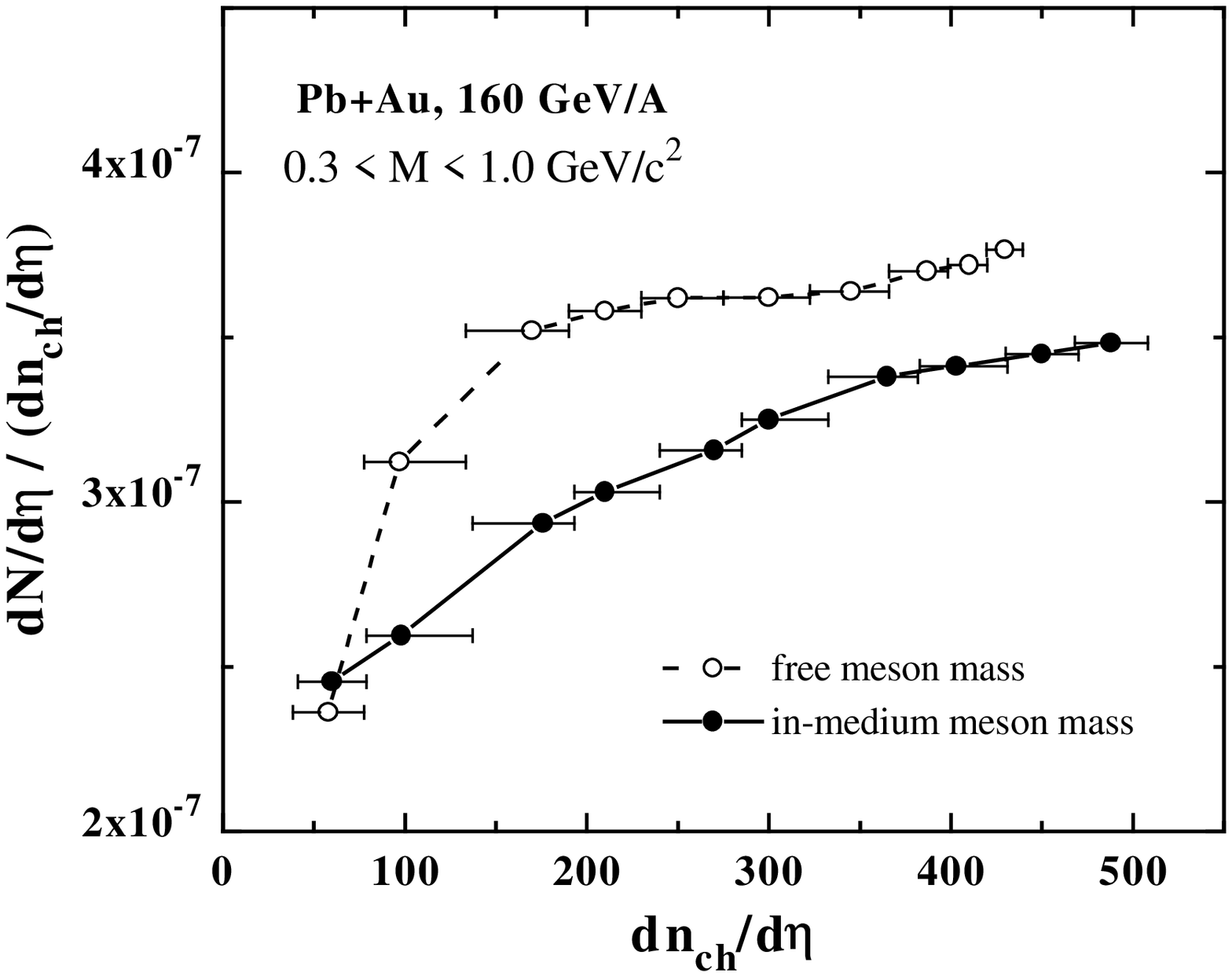,width=15cm,height=18cm}}
\vspace*{-5.3cm}
\caption{The differential dilepton spectra for Pb~+~Au at 160~GeV/u
integrated over the invariant mass region $0.3 \le M \le 1.0$~GeV as a
function of the charged particle  multiplicity $dn_{ch}/d\eta$
without (open circles) and with (solid circles) the in-medium mass
modification of the mesons.}
\label{Fig7}
\end{figure}

The number of charged particles $dn_{ch}/d\eta$ in the pseudorapidity
bin 2.1 $\leq \eta \leq$ 3.1 for Pb~+~Au at 160~GeV/u is shown in
Fig.~\ref{Fig6}.  The open circles are the result of our computations
with 'free' meson masses, while the solid circles correspond to
calculations when including the in-medium modifications of the meson
masses.  In both cases the charged particle multiplicity decreases with
impact parameter practically linearly. For peripheral collisions there
is no essential difference between both schemes as expected. For
central collisions, however, the charged particle multiplicity in the
'dropping' meson mass scenario is slightly larger due to the reduction
of the vector meson and antikaon  production thresholds, which enhances
the respective particle formation cross sections at high baryon
density. Especially the subsequent decay of $\rho$ and $\omega$ mesons
to pions leads to a larger number of pions in the final expansion
phase. We note, that due to the conservation of energy and momentum in
each production event the enhanced number of vector mesons and
antikaons at finite baryon density goes along with a lower number of
those mesons, that do not change their quasiparticle properties in the
medium ($\pi, \eta, \eta^\prime $ etc.).

Including the CERES acceptance cuts and mass resolution as described above,
we show in Fig.~\ref{Fig7} the dilepton yield
integrated over the invariant mass range $0.3 \le M \le 1.0$~GeV,
\begin{eqnarray}
{dN/d\eta \over dn_{ch}/d\eta} =
\int\limits_{0.3\ {\rm GeV}}^{1.0\ {\rm GeV}} dM \ {dn_{e^+e^-}/(dM d\eta)
\over dn_{ch}/d\eta}
\label{intdndy}\end{eqnarray}
as a function of the charged particle multiplicity $dn_{ch}/d\eta$
without (open circles) and with (solid circles) in-medium mass
modification.  At small charged particle multiplicity, which
corresponds to very peripheral collisions, the integrated dilepton
yields coincide for both cases. With decreasing impact parameter the
average baryon density and especially the pion density increases; as a
consequence the contribution from pion annihilation to $\rho^0$ and
subsequent decay to dileptons becomes larger.  Since we gate on
dileptons above the $\pi^+\pi^-$ annihilation threshold, also the
integrated dilepton spectra increase with $dn_{ch}/d\eta$. Using 'free'
meson masses we reach some plateau for low impact parameter which
implies that the $\pi \pi$ annihilation contribution, divided by the
pion density, becomes approximately constant. However, for the
'dropping' mass scenario the absolute dilepton yield above 0.3~GeV is
smaller because the directly produced $\rho$-mesons (at high initial
baryon density) also 'shine' in the invariant mass regime below 0.3~GeV
(cf. lower part of Fig.~\ref{Fig5}).  Furthermore, due to an initially
higher vector meson density the initial pion density (due to energy
conservation) is reduced as compared to the 'free' mass scenario and
the corresponding pion annihilation contribution is also lowered to
some extend.  All effects together lead to an approximately
linear increase of the integrated dilepton yield with the charged
particle multiplicity in the 'dropping' mass picture. Experimental data
with sufficient statistics should allow to disentangle the two schemes
or disqualify the hadronic scenario as employed in the HSD transport
approach.

\section{Systematics of dilepton production from AGS to SPS energies}

In this section we present a systematic analysis of various
dilepton observables -- the differential dilepton spectra, rapidity and
transverse momentum distributions, the average transverse momentum -- for
p~+~Be and Pb~+~Pb collisions from AGS to SPS energies for the 'free' and
'dropping' mass scenarios.

\subsection{Differential dilepton spectra}

\begin{figure}
\vspace*{-3cm}
{\psfig{figure=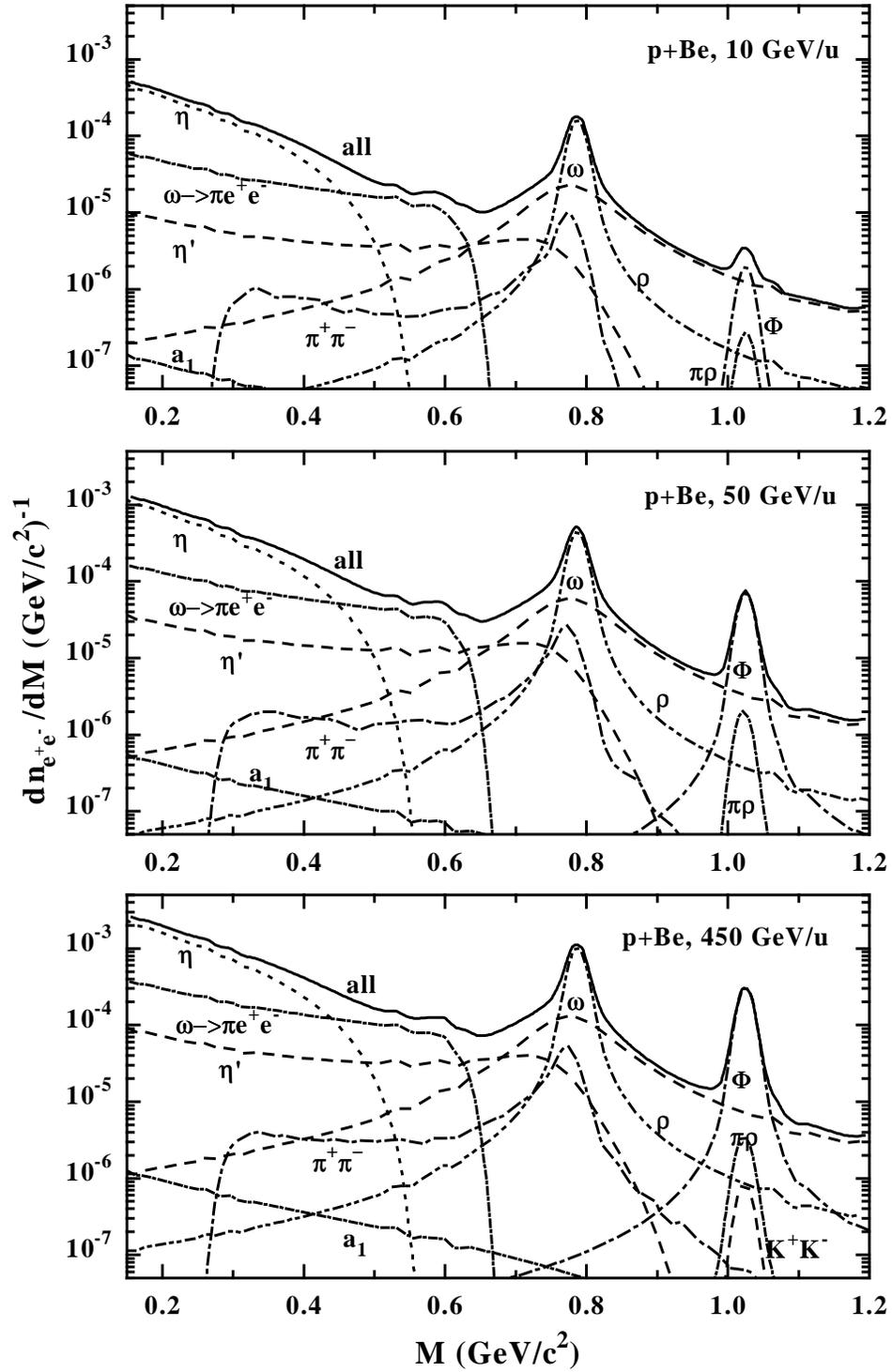,width=15cm,height=22cm}}
\vspace*{-.5cm}
\caption{The differential dilepton spectra for p~+~Be at 10, 50, and 450~GeV.
The thick solid lines display the sum of all channels whereas the individual
contributions are given in terms of the thinner lines.
The assignment of
the individual contributions is the same as in Fig. 4. The mass resolution
employed is $\Delta M$ = 10 MeV.}
\label{Fig8}
\end{figure}

\begin{figure}
\vspace*{-3cm}
{\psfig{figure=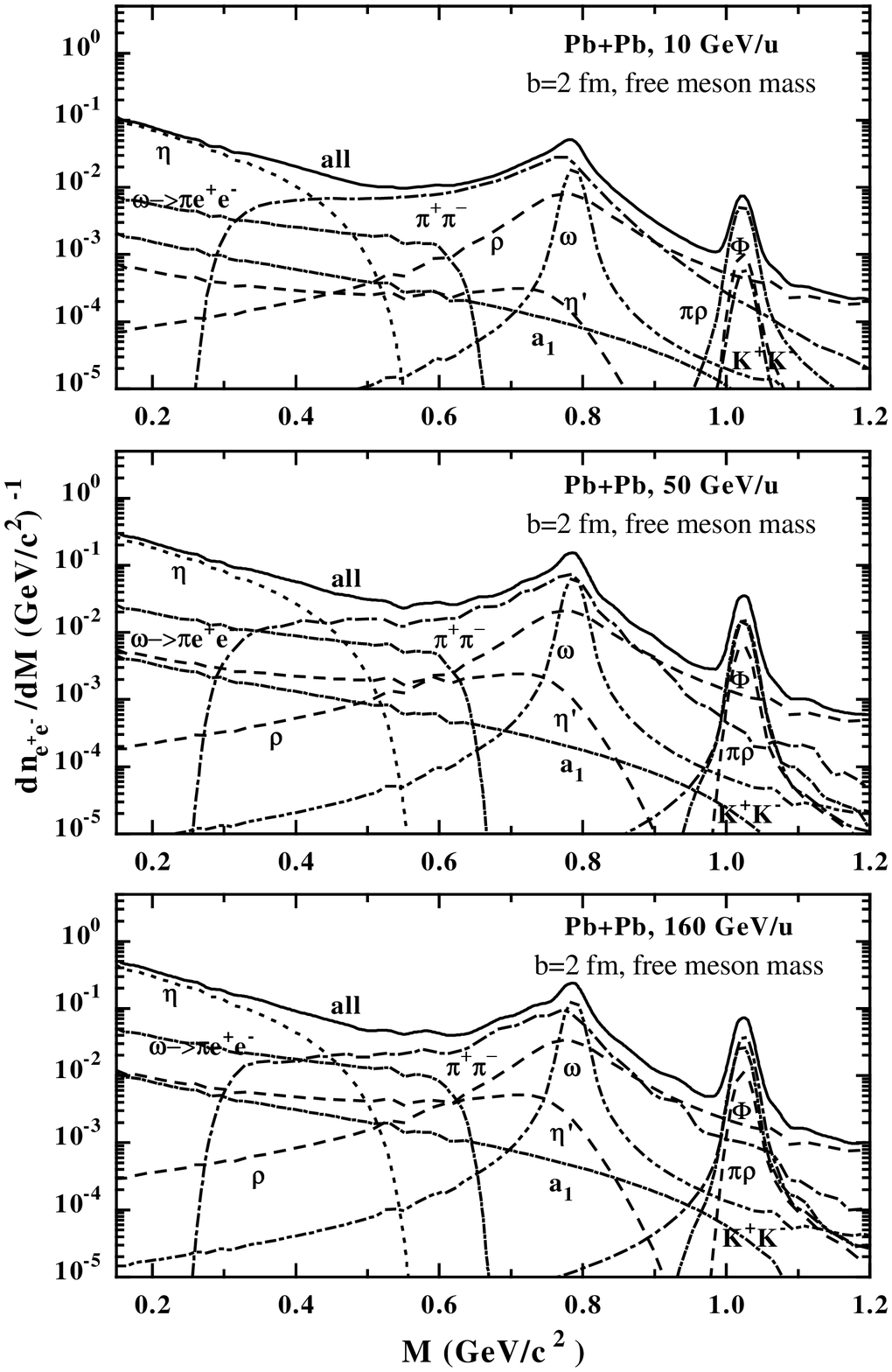,width=15cm,height=22cm}}
\vspace*{-.5cm}
\caption{The differential dilepton spectra for Pb~+~Pb at 10, 50, 160~GeV/u
at $b=2$~fm within the 'free' meson mass scenario. The assignment of
the individual contributions is the same as in Fig. 4. The mass resolution
employed is $\Delta M$ = 10 MeV.}
\label{Fig9}
\end{figure}

\begin{figure}
\vspace*{-3cm}
{\psfig{figure=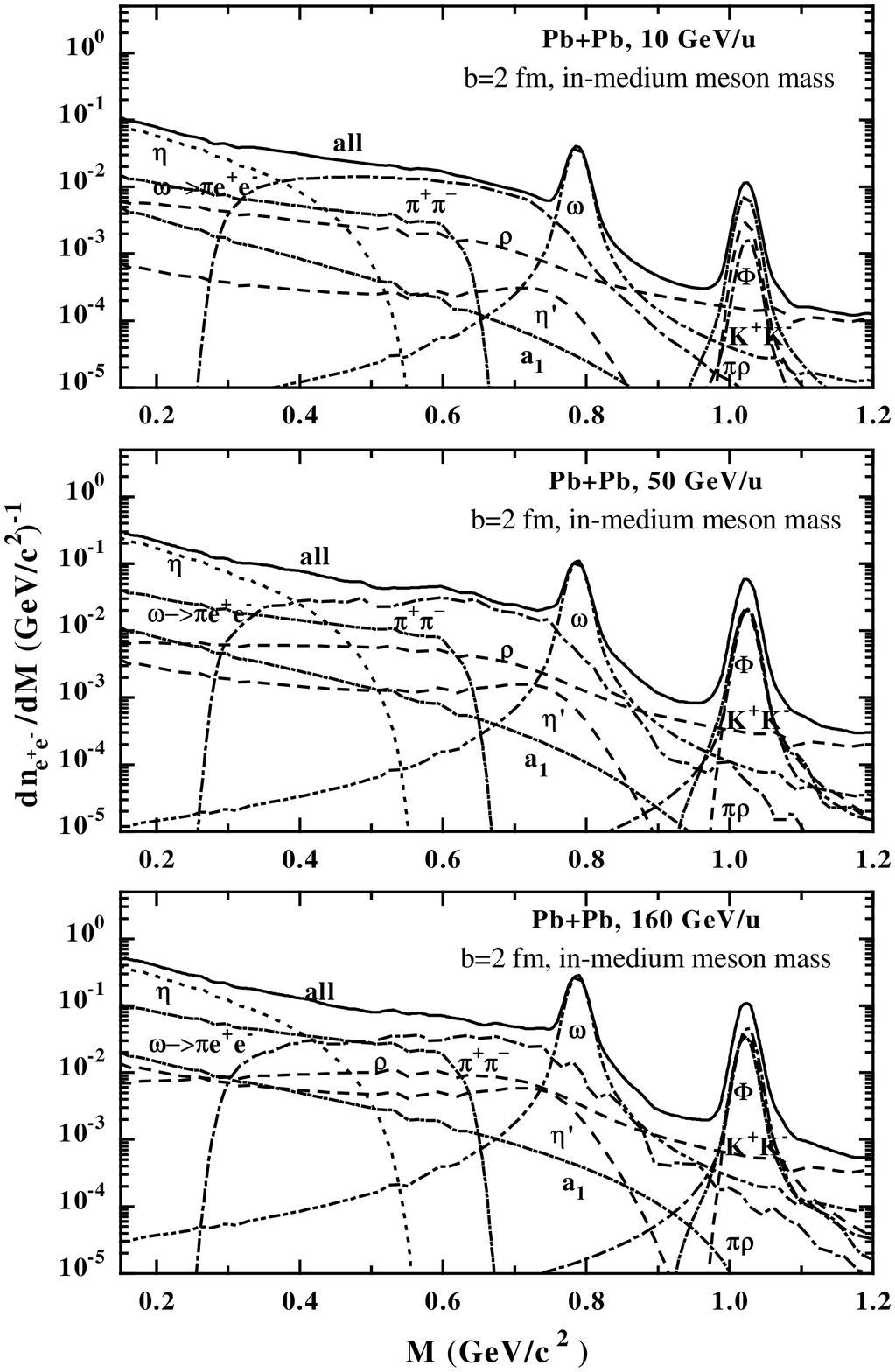,width=15cm,height=22cm}}
\vspace*{-.5cm}
\caption{The differential dilepton spectra for Pb~+~Pb at 10, 50, 160~GeV/u
at $b=2$~fm within the in-medium meson mass scenario.
The assignment of
the individual contributions is the same as in Fig. 4. The mass resolution
employed is $\Delta M$ = 10 MeV.}
\label{Fig10}
\end{figure}

We first examine the differential dilepton spectra - integrated over
rapidity and transverse momentum - with respect to their 'cocktail'
decomposition.  The respective spectra for p~+~Be at 10, 50, 450~GeV/u
and for Pb~+~Pb at 10, 50, 160~GeV/u for $b=2$~fm  are shown in
Figs.~\ref{Fig8} -- \ref{Fig10}. For all cases we include a mass
resolution of $\Delta M = 10$~MeV, which can be expected for future
dilepton detector systems \cite{Johanna}. As seen from
Figs.~\ref{Fig8} -- \ref{Fig10}, there are no dramatic differences in the
relative contribution of the various dilepton channels; the total yield
(and especially the $\Phi$ decay) increase with bombarding energy quite
smoothly.  In case of central Pb~+~Pb collisions the dominance of the
$\pi^+ \pi^-$ annihilation component is most pronounced at 10~GeV/u in
both scenarios and decreases with bombarding energy.

\begin{figure}
\vspace*{-3cm}
{\psfig{figure=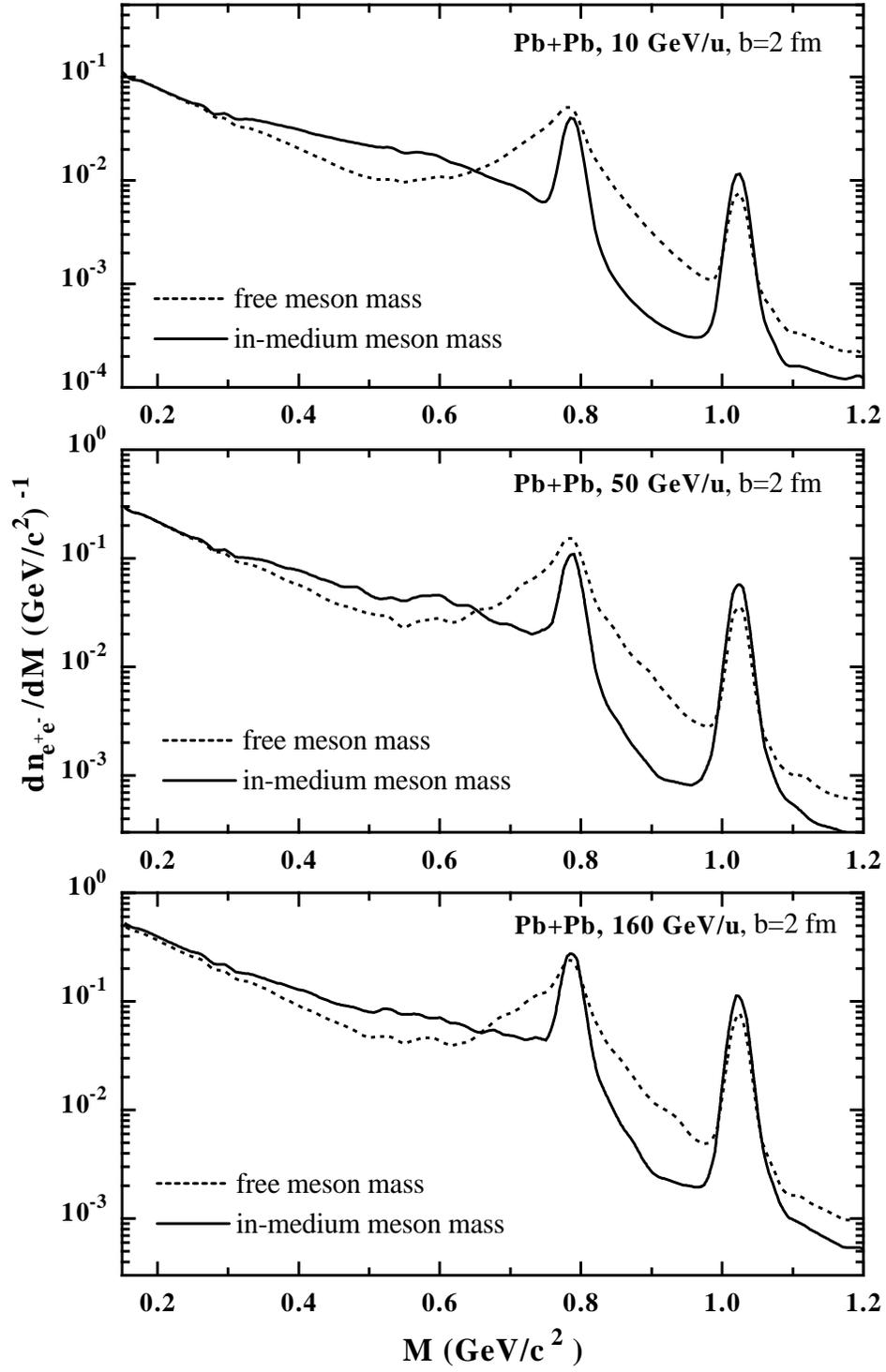,width=15cm,height=22cm}}
\vspace*{-.5cm}
\caption{The differential dilepton spectra for Pb~+~Pb
at 10, 50, 160~GeV/u at $b=2$~fm without (dashed lines)
and with in-medium meson mass modification (solid lines) for a mass
resolution $\Delta M$ = 10 MeV.}
\label{Fig11}
\end{figure}

Since experimentally only the total spectra can be observed, we show in
Fig.~\ref{Fig11} the sum of all contributions for the 'free' (dashed
lines) and 'dropping' mass scenario (solid lines) for central
collisions of Pb~+~Pb at different bombarding energies to demonstrate
the influence of in-medium effects for the mesons.  For a mass
resolution $\Delta M = 10$~MeV one can expect to observe not only the
"usual" enhancement of the dilepton spectra by about a factor of 2 at
invariant masses $0.3 \le M \le 0.6$~GeV, but also a sharp drop of the
spectrum above the $\omega$ mass by a factor 4 -- 6 due to the shift of
the $\rho$ contribution to lower invariant masses. Furthermore, the
peak from the $\omega$ meson becomes more pronounced since the
'background' from the $\rho$ decay (either from $\pi B$, $BB$ or $\pi
\pi$ collisions) is significantly reduced. In the $\Phi$ mass region,
furthermore, we find a small increase of the yield for the 'dropping'
mass scheme at all energies from 10 -- 160~GeV/u. Again, the relative
modifications of the total spectrum are most pronounced at 10~GeV/u.

\subsection{Transverse momentum distributions}

\begin{figure}
\vspace*{-3cm}
{\psfig{figure=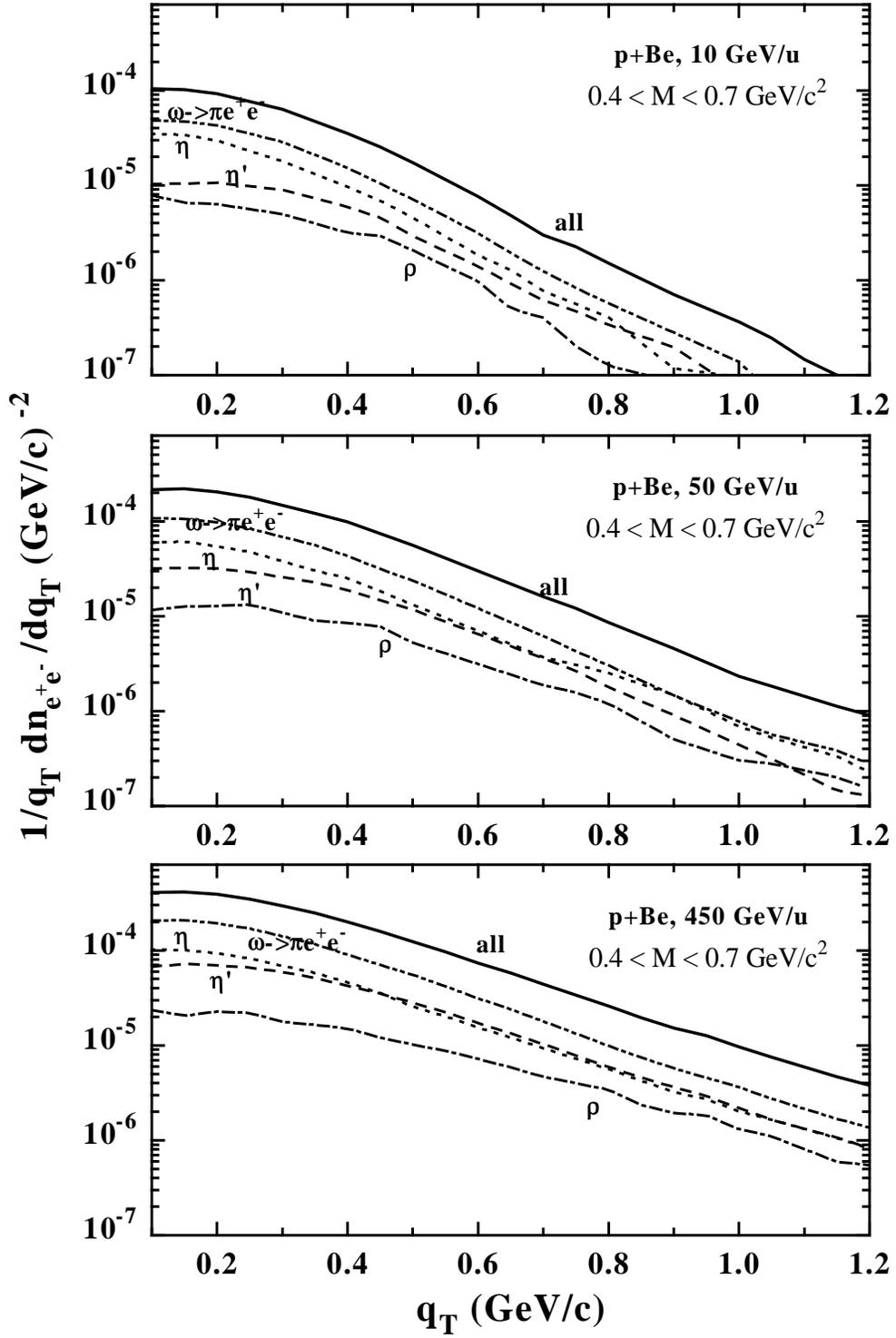,width=15cm,height=22cm}}
\vspace*{-.5cm}
\caption{The channel decomposition for the average transverse momentum
distribution in the invariant mass range $0.4 \le M \le 0.7$~GeV
for p~+~Be at 10, 50, and 450~GeV;
all contributions (solid line), $\omega \to \pi^0 e^+e^-$ (dot-dot-dashed line),
$\eta \to \gamma e^+e^-$ (short dashed line), $\eta^\prime \to \gamma e^+e^-$
(long dashed line), $\rho^0 \to e^+e^-$ (dot-dashed line).}
\label{Fig12}
\end{figure}

\begin{figure}
\vspace*{-3cm}
{\psfig{figure=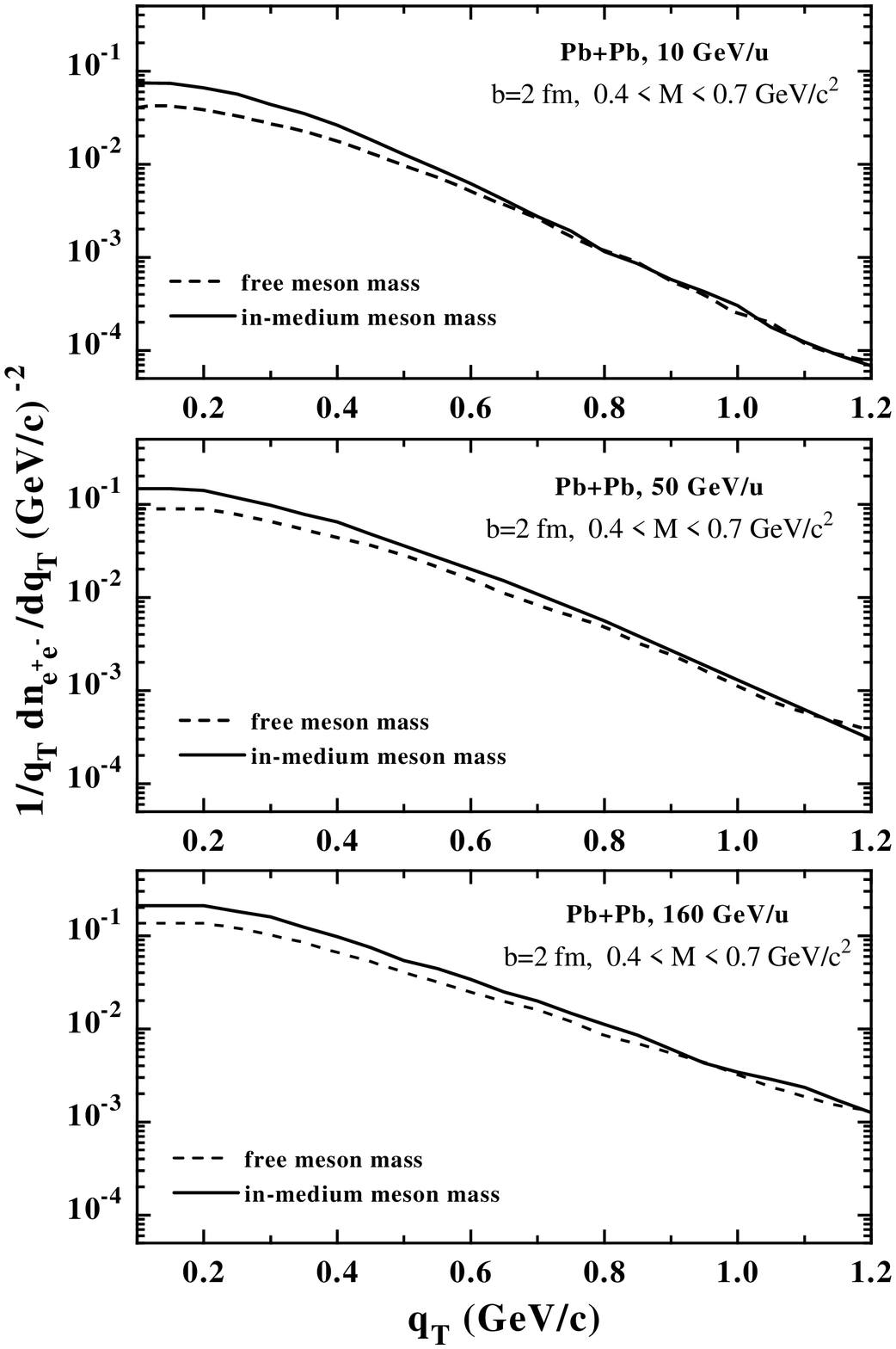,width=15cm,height=22cm}}
\vspace*{-.5cm}
\caption{The transverse momentum distribution for the invariant mass
range $0.4 \le M \le 0.7$~GeV for Pb~+~Pb at 10, 50, and 160~GeV/u
at $b=2$~fm without (dashed lines) and with in-medium meson mass
modification (solid lines).}
\label{Fig13}
\end{figure}

\begin{figure}
\vspace*{-2cm}
{\psfig{figure=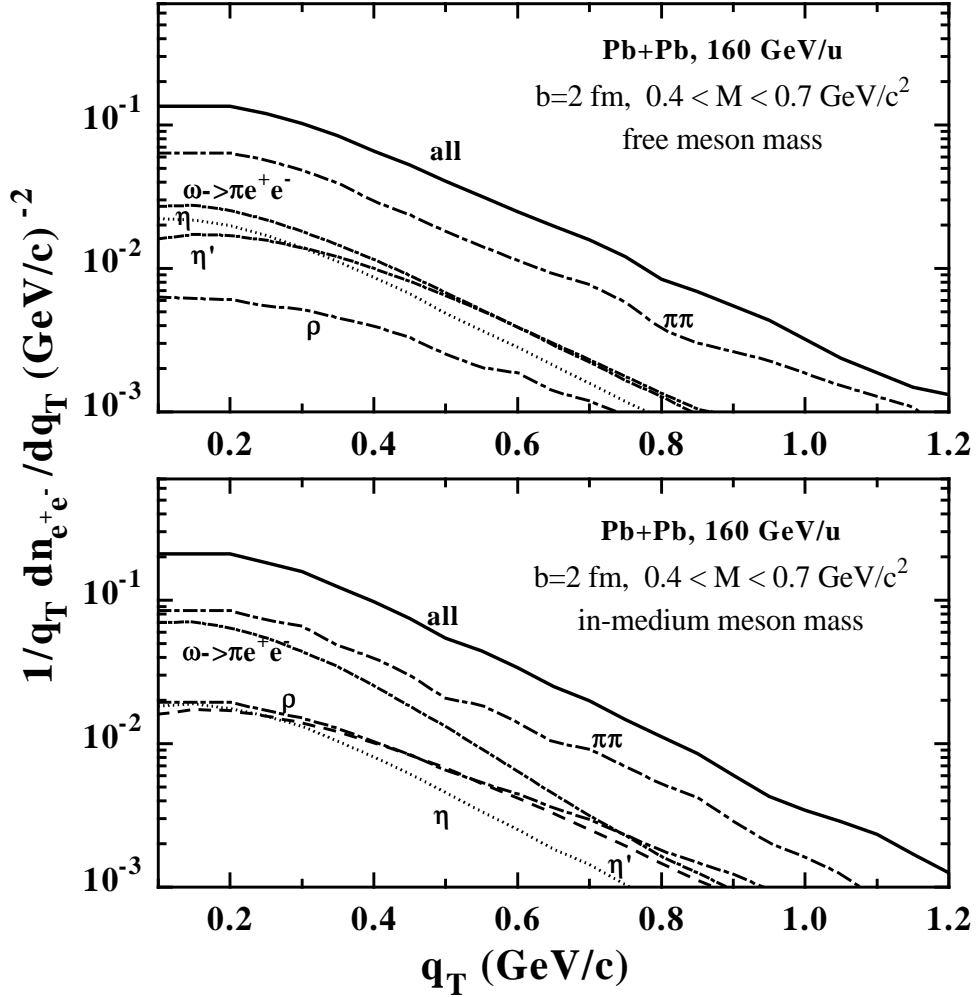,width=15cm,height=22cm}}
\vspace*{-2cm}
\caption{The channel decomposition for the transverse momentum distribution
for the invariant mass range $0.4 \le M \le 0.7$~GeV for Pb~+~Pb at
160~GeV/u at $b=2$~fm without (upper part) and with in-medium meson mass
modification (lower part);
all contributions (solid line), $\pi^+ \pi^- \to e^+e^-$ (upper dot-dashed 
line),
$\omega \to \pi^0 e^+e^-$ (short dot-dashed line),
$\eta \to \gamma e^+e^-$ (dotted line), $\eta^\prime \to \gamma e^+e^-$
(dashed line), $\rho^0 \to e^+e^-$ (lower dot-dashed line).}
\label{Fig14}
\end{figure}

In this subsection we explore if the transverse momentum distribution
of the lepton pair might give some further criteria to distinguish
between the different scenarios. In this respect the transverse
momentum distributions -- integrated over rapidity and the invariant
mass $0.4 \le M \le 0.7$~GeV --  are shown in
Figs.~\ref{Fig12} -- \ref{Fig14} for p~+~Be at 10, 50, 450~GeV/u and for
Pb~+~Pb at 10, 50, 160~GeV/u for $b=2$~fm using $q_\perp$-bins of 50~MeV/c.

The main contribution in the invariant mass region  $0.4 \le M \le
0.7$~GeV for p~+~Be comes from the $\eta,\omega$ Dalitz decays
according to Fig.~\ref{Fig8}. The same decomposition can be observed in
Fig.~\ref{Fig12}, where we display only the dominant channels.
Performing an exponential fit to the sum of all contributions (thick
solid lines in Fig.~\ref{Fig12}) for  $M\geq 0.4$~GeV the following
slope parameters can be extracted from our calculations:  125~MeV at
10~GeV/u, 160~MeV at 50~GeV/u, and 185~MeV at 450~GeV/u.

Contrary to p~+~Be reactions, the dominant channels in the mass range
$0.4 \le M \le 0.7$~GeV for central Pb~+~Pb collisions are the pion
annihilation, $\omega$ Dalitz decay and direct decay of vector mesons.
As shown before, including the in-medium 'dropping' masses we obtain a
shift of the $\rho$ meson contribution (cf.
Figs.~\ref{Fig9},\ref{Fig10}) to smaller invariant masses and one might
expect same enhancement of the transverse momentum distribution at
lower $q_\perp$ in this case.  This effect is demonstrated  in
Fig.~\ref{Fig13} where the transverse momentum spectra for the
in-medium mass scheme (solid lines) are slightly softer than for the
'free' mass case (dashed lines). This tendency can also be seen from
the slope parameters extracted for $q_T \geq 0.4$~GeV/c:  145~MeV at
10~GeV/u for the  'free' meson mass scheme -- 130~MeV, when including
the in-medium mass modification; at 50~GeV/u -- 170~MeV and 155~MeV,
respectively; at 160~GeV/u -- 195~MeV and 180~MeV, respectively.

In order to clarify the origin of these differences, we present
the channel decomposition for central Pb~+~Pb collisions at 160~GeV/u
in Fig.~\ref{Fig14} for both scenarios. The pion annihilation
and $\omega$ Dalitz decay contribution, which
are dominant in the invariant mass region considered, become slightly softer
in the 'dropping' mass scenario. We note, however, that the transverse
momentum spectra do not differ sizeably; according to the authors point
of view they do not qualify very much for disentangling the different schemes.

\subsection{Average transverse momentum $<q_T>(M)$}

The dilepton average transverse momentum as a function of the
invariant mass, \mbox{$<q_T>(M)$}, was already
studied experimentally for hadron-hadron collisions
a couple of years ago (cf. the reviews
\cite{Craigie78,Stroyn81}), where an increase of $<q_T>$ with
M and bombarding energy was observed.

\begin{figure}
\vspace*{-3cm}
{\psfig{figure=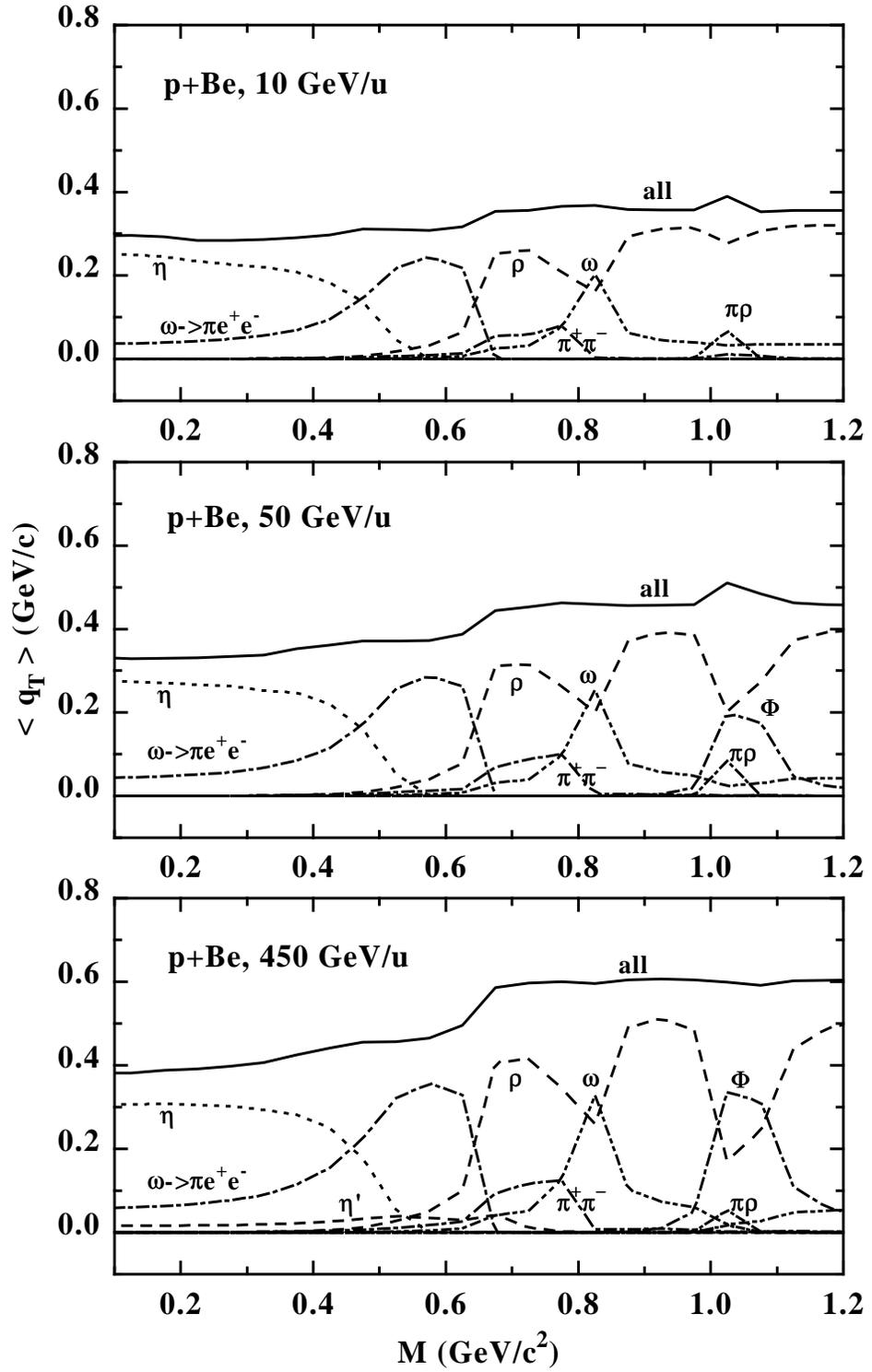,width=15cm,height=22cm}}
\vspace*{-.5cm}
\caption{The average transverse momentum $<q_T>(M)$
including the channel decomposition for p~+~Be at 10, 50 and 450~GeV.
The assignment of
the individual contributions is the same as in Fig. 4.}
\label{Fig15}
\end{figure}

In our analysis the average transverse momentum for the
channel $k$ is defined as
\begin{eqnarray}
<q_T>^k(M) = {\sum\limits_i w_i^k({q_T}_i, M_i) \ {q_T}_i \over
\sum\limits_k \sum\limits_i w_i^k({q_T}_i, M_i)},
\label{qtav}\end{eqnarray}
where $w_i^k({q_T}_i, M)$ is the probability of the individual dilepton
event $i$ (cf. Section 2.2). The total $<q_T>$ is the sum of the
individual channels:
\begin{eqnarray}
<q_T>(M) = \sum\limits_k <q_T>^k(M).
\label{qtavtot}\end{eqnarray}
The results of our computations are shown in Fig.~\ref{Fig15} for p~+~Be
at 10, 50 and 450~GeV  including the contributions from the
individual channels.  Due to mass bins of 50~MeV the shapes from the
vector mesons $\omega$ and $\Phi$ are not very pronounced.  However,
one can see the main tendency: the total $<q_T>$ increases with
invariant mass and bombarding energy in the same way as for
hadron-hadron collisions \cite{Craigie78,Stroyn81}.

\begin{figure}[h]
\vspace*{-3cm}
{\psfig{figure=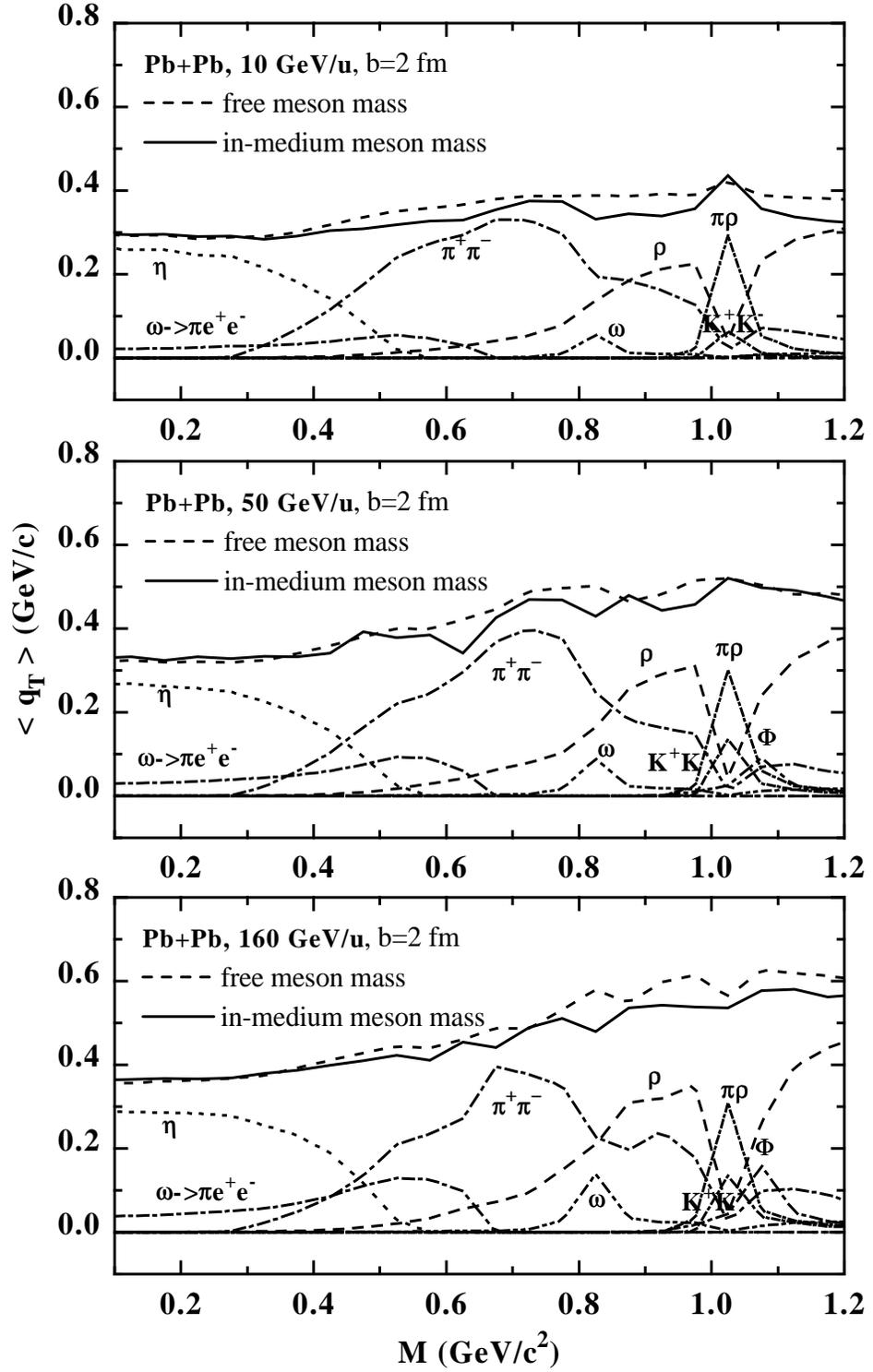,width=15cm,height=22cm}}
\vspace*{-.5cm}
\caption{The average transverse momentum $<q_T>(M)$
including the channel decomposition for Pb~+~Pb at 10, 50 and 160~GeV/u
at $b=2$~fm without (dashed lines) and with in-medium meson mass
modification (solid lines). The assignment of
the individual contributions is the same as in Fig. 4.}
\label{Fig16}
\end{figure}

A similar behaviour of the average transverse momentum is found for
central collisions of Pb~+~Pb at 10, 50 and 160~GeV/u in
Fig.~\ref{Fig16}, were we have plotted the total $<q_T>$ for both
scenarios, i.e. without (dashed lines) and with (solid lines)
'dropping' of meson masses. The channel decomposition is shown only for
the 'free' mass scenario. Contrary to p~+~Be the contribution from
$\pi^+\pi^-$ annihilation dominates from 0.5 -- 0.8~GeV at all bombarding
energies, however, we do not find a pronounced difference in the total
$<q_T>$ between both mass schemes.

\subsection{Rapidity distribution of lepton pairs}

\begin{figure}
\vspace*{-3cm}
{\psfig{figure=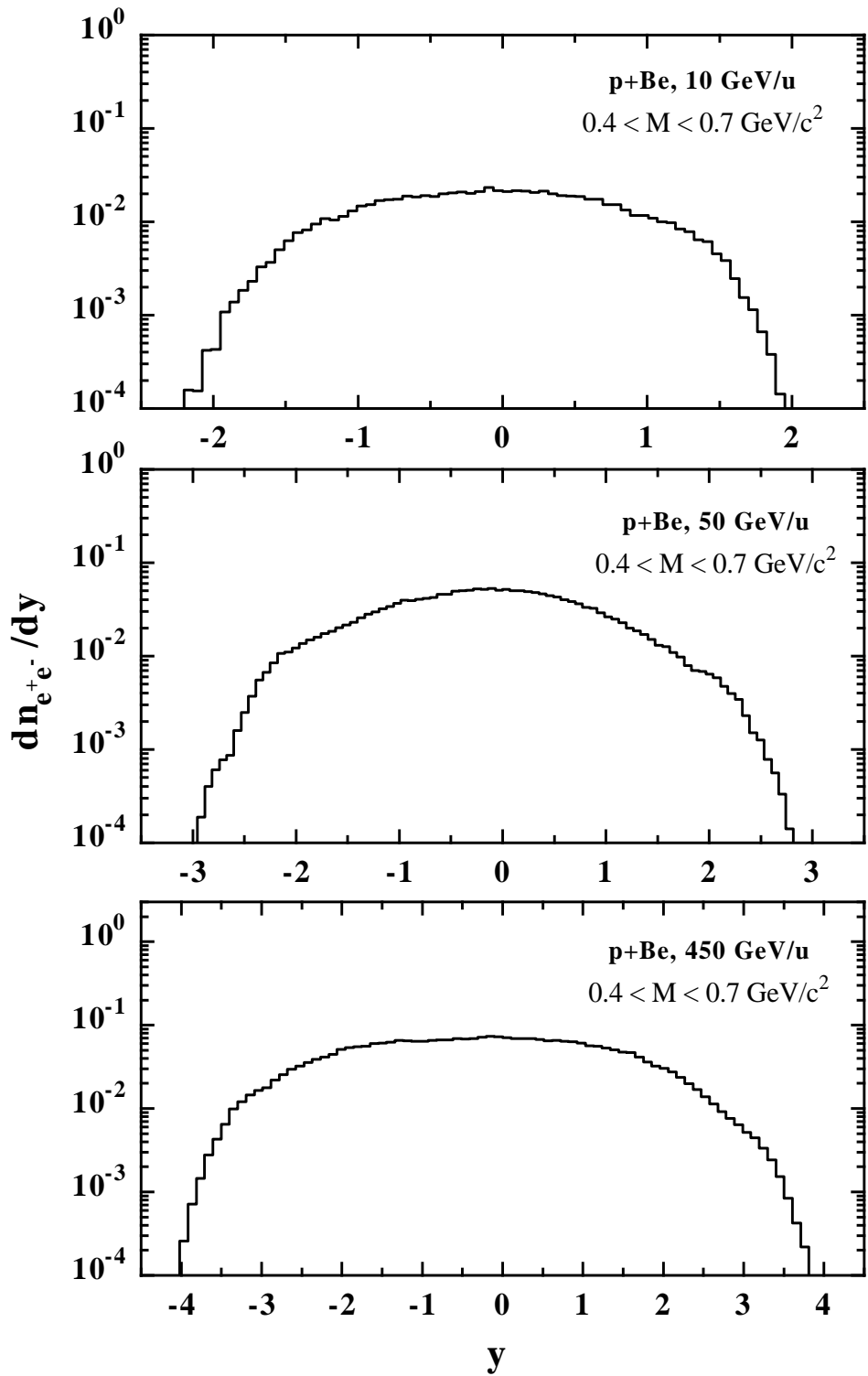,width=15cm,height=22cm}}
\vspace*{-.5cm}
\caption{The dilepton rapidity distribution for the invariant mass range
$0.4 \le M \le 0.7$~GeV for p~+~Be at 10, 50, and 450~GeV.}
\label{Fig17}
\end{figure}

\begin{figure}
\vspace*{-3cm}
{\psfig{figure=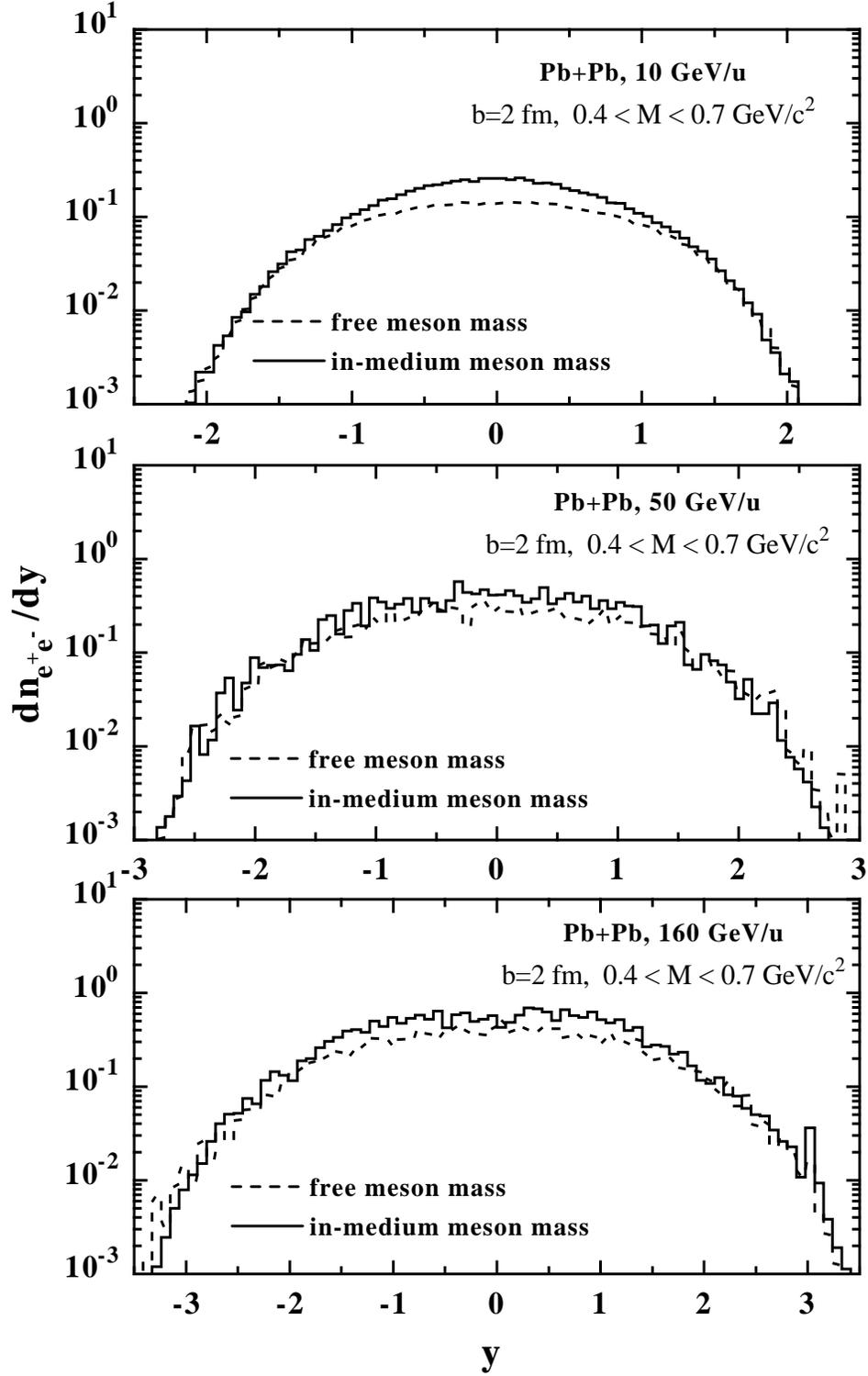,width=15cm,height=22cm}}
\vspace*{-.5cm}
\caption{The dilepton rapidity distribution for the invariant mass range
$0.4 \le M \le 0.7$~GeV for Pb~+~Pb at 10, 50 and 160~GeV/u at $b=2$~fm
without (dashed lines) and with in-medium meson mass modification
(solid lines).}
\label{Fig18}
\end{figure}

The results of our calculations for the dilepton rapidity distribution
in the nucleon-nucleon center-of-mass system are displayed in
Figs.~\ref{Fig17},\ref{Fig18}.  In analogy to the previous cases we
study the systems  p~+~Be at 10, 50, 450~GeV/u and Pb~+~Pb at 10, 50,
160~GeV/u  ($b=2$~fm) integrating over the transverse momentum and the
invariant mass range $0.4 \le M \le 0.7$~GeV.  As seen from
Figs.~\ref{Fig17},\ref{Fig18} the shapes of the rapidity distributions
are quite similar.  An enhancement of about a factor 1.5 -- 2 for Pb~+~Pb
can be seen at midrapidity when employing in-medium meson masses (solid
lines), while there is practically no difference at more forward or
backward rapidities.  Again this enhancement is most pronounced at
10~GeV/u for Pb~+~Pb and decreases with bombarding energy as expected
from Fig.~\ref{Fig1}.

\section{Direct photons}

Directly radiated thermal photons have been considered as an
independent probe to study the hot and dense nuclear matter produced in
ultrarelativistic nucleus-nucleus collisions \cite{Ruuskanen,Kapusta}.
However, an experimental measurement of direct photons is a quite
complicated task due to the background from hadronic decays. Only
recently first upper limits for direct photon spectra have been
reported by the WA80 collaboration~\cite{WA80} for S~+~Au at 200~GeV/u.
In the latter study photons from $\pi^0$ and $\eta$ Dalitz decays have
been subtracted from the total photon signal; their spectra thus can be
interpreted as upper bounds for the direct photon cross section.

A first calculation of the direct photon radiation from a
quark-gluon-plasma (QGP) was performed a couple of years ago in
Ref.~\cite{Shuryak}; various hydrodynamical model calculations followed
(cf.~\cite{WA80,Shur96} and references therein), where the radiation
from a QGP \cite{Sivastava} has been compared to the radiation from a
pure hadron gas scenario \cite{Dumitru}.  The comparison of the various
models with the WA80 data \cite{WA80}, however, has demonstrated only
the inapplicability of hadronic thermal models with high initial
temperature.  In this respect it should be quite useful to compare the
WA80 upper limits with the results of a nonthermal model -- such as the
HSD transport approach -- to find out possible conflicts with the hadronic
scenario employed.

\subsection{Description of elementary channels}

In our analysis we take into account the following processes for photon
production: $a_1 \to \pi\gamma$, $\omega \to \pi\gamma$, $\eta^\prime
\to \rho\gamma$ or $\omega \gamma$. The $\pi^0$ and $\eta$ decays are
already subtracted experimentally and thus don't have to be taken into
account in our calculations.

The treatment of photon production in the HSD approach is quite similar
to that for dileptons, however, using  the branching ratios:
\begin{eqnarray}
{\Gamma_{a_1\to \pi\gamma}\over \Gamma_{a_1}} \simeq 1.6\times 10^{-3},\
{\Gamma_{\omega\to \pi\gamma}\over \Gamma_{\omega}} = 0.085,  \
{\Gamma_{\eta^\prime\to \rho\gamma}\over \Gamma_{\eta^\prime}} = 0.3, \
{\Gamma_{\eta^\prime\to \omega\gamma}\over \Gamma_{\eta^\prime}} = 0.03.
\label{photon}\end{eqnarray}
We discard baryon-baryon, meson-baryon and meson-meson bremsstrahlung
in our present study since these channels were found to be of minor
importance for dilepton production in case of S~+~Au at 200~GeV/u in
Ref.~\cite{Cass95C}.  Furthermore, the soft-photon approximation
employed in \cite{Cass95C} is questionable at these energies according
to the studies in Refs.~\cite{Lichard95,Eggers}.

\subsection{Comparison with experimental data for S~+~Au}

\begin{figure}
\vspace*{-2cm}
{\psfig{figure=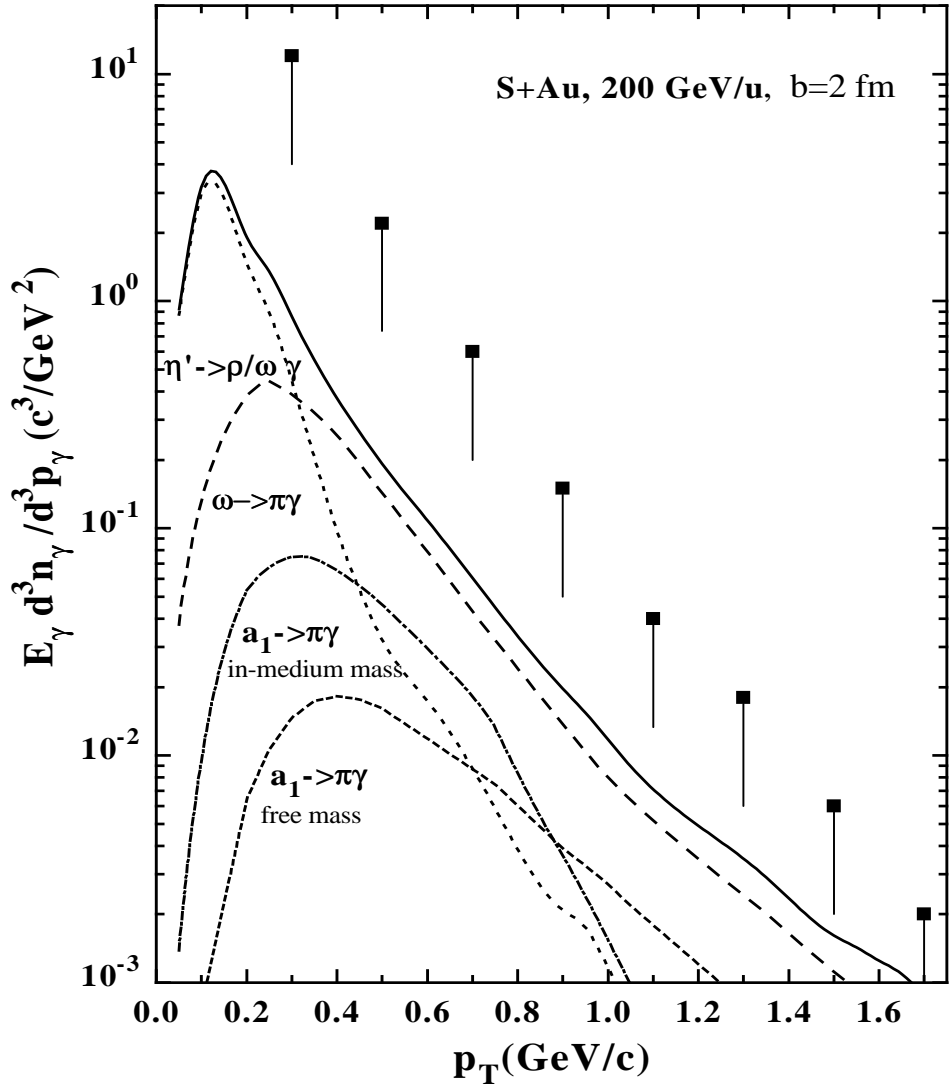,width=15cm,height=20cm}}
\vspace*{-2cm}
\caption{The differential photon multiplicity for central S~+~Au collisions
at 200~GeV/u in comparison to the upper limits from the WA80 collaboration
\protect\cite{WA80}; all contributions (solid line).}
\label{Fig19}
\end{figure}

In Fig.~\ref{Fig19} we show the result of our calculations for photon
production in central S~+~Au collisions at 200~GeV/u in comparison with
the experimental data~\cite{WA80}.  The computations were performed at
$b=2$~fm including the experimental rapidity cut $2.1 \le y \le 2.9$.
As seen form Fig.~\ref{Fig19} the main contribution in our calculation
comes from $\eta^\prime \to \rho/\omega \ \gamma$ decays at low $q_T
\le 0.4$~GeV and from $\omega\to \pi\gamma$ for $q_T \ge 0.4$; the
solid line is the sum of all contributions which is still well below
the upper limits of WA80.

For the process $a_1\to \pi\gamma$ we explore again both scenarios,
i.e.  without (dashed line) and with (dashed-dotted line) in-medium
mass modification.  In fact, the dropping of the $a_1$ mass leads to a
sizeable enhancement of $a_1$ mesons in the reaction zone and thus to a
significant enhancement of the photon spectra from the $a_1$ as pointed
out in Refs.~\cite{Xiong92,Shur96}.  However, the relative contribution
from the  $a_1$ decay is still far below the 'background' from
$\eta^\prime$ and $\omega$ decays. Thus, even in case of the 'dropping'
meson masses we do not get in any conflict with the upper limits
imposed by the WA80 data \cite{WA80}.

\section{Summary}

In this work we have studied dilepton production in proton and
heavy-ion induced reactions from 10 -- 450~GeV or 10 -- 160~GeV/u,
respectively, on the basis of the covariant transport approach
HSD~\cite{Ehehalt}.  We have incorporated the contributions of the
Dalitz-decay of the $\eta, \omega, \eta^\prime, a_1$ as well as
$\pi^+\pi^-$ annihilation and the direct dilepton decay of the vector
mesons $\rho, \omega, \Phi$ as well as $K^+K^-$ and $\pi \rho$
channels.  It is found that for p~+~Be at 450~GeV the mesonic decays
almost completely determine the dilepton yield, whereas in Pb~+~Au
reactions the $\pi^+\pi^-, K^+ K^-$ annihilation channels and $\pi
\rho$ collisions contribute substantially.  The experimental data taken
by the CERES collaboration~\cite{CERES,Ullrich} generally are
underestimated by the calculations for invariant masses 0.35~GeV~$\leq
M \leq 0.65$~GeV when using 'free' form factors for the pion and
$\rho$-meson in line with Refs.~\cite{Li,Koch96,Li96}.

We have, furthermore, examined the effects of 'dropping' $\rho,
\omega, \Phi, K^-$ and $a_1$ masses on the dilepton spectra, which
generally leads to an improvement in the description of the experimental
data from the CERES~\cite{CERES,Ullrich} and HELIOS-3 collaborations
(cf. also Refs.~\cite{Li,Cass95C,Cass96H,Li96}). In order to allow for
a clearer distinction between the different scanarios we have performed
systematic studies for p~+~Be and central Pb~+~Pb collisions from 10  --
450 or 160~GeV/u, respectively, with respect to the transverse momentum
distribution of the dilepton pair and their rapidity distribution. At
all bombarding energies the in-medium effects are most pronounced at
midrapidity, but the modifications in the transverse momentum spectra
are only very moderate, such that the $q_\perp$ distributions do not
qualify very much for quantifying the in-medium effects. The same holds
for the average transverse momentum of the dilepton pair as a function
of the invariant mass $M$. A more pronounced variation is obtained for
the dilepton yield (integrated for $M \geq$ 0.3~GeV) as a function of
the charged particle multiplicity at midrapidity (cf. Fig.~\ref{Fig7}),
where the 'free' and 'dropping' meson mass scenarios differ
significantly. This sensitivity should be explored in the next round of
experiments at the SPS.

Apart from the enhancement of the $e^+e^-$ yield for invariant masses
$0.3 \leq M \leq 0.7$~GeV in case of 'dropping' meson masses, one
should also see a decrease by factors 4 -- 6 in the latter scheme for
$0.8 \leq M \leq 1$~GeV in case of central Pb~+~Pb collisions (cf.
Fig.~\ref{Fig11}), provided that the experimental mass resolution is in
the order of about 10~MeV. The relative differences between the 'free'
and 'dropping' mass schemes become larger when decreasing the
bombarding energy from 160~GeV/u to 50~GeV/u and even to 10~GeV/u.
According to Fig.~\ref{Fig1} there might be a maximum sensitivity
around 20 -- 30~GeV/u, however, the relative change of the dilepton
spectrum from 10 -- 50~GeV/u -- apart from an overall increase by about a
factor of 3 -- is not very pronounced.

We have, furthermore, explored if the hadronic reactions rates from the
HSD transport approach might come in conflict with the upper limits for
the direct photon spectrum in central collisions of S~+~Au from the
WA80 collaboration \cite{WA80}, because the photon multiplicity from
$a_1$ decay might be dramatically enhanced in a phase with a partial
restoration of chiral symmetry as suggested in Ref.~\cite{Shur96}. Our
computations, however, including also the processes $\eta^\prime \to
\rho/\omega \ \gamma; \omega \to \pi \gamma$ do not show any
indications for this, since the sum-spectra are still below the upper
limits from WA80 by at least a factor of 3. In this respect we do not
find any inconsistencies within the hadronic picture of the
nucleus-nucleus collision. As pointed out in Ref.~\cite{Casko}, there
might be problems due to the large energy densities reached in central
Pb~+~Pb collisions (above 3~GeV/fm$^3$) at 160~GeV/u, where one
generally believes that droplets of a quark-gluon-plasma (QGP) should
be formed.

We finally note, that apart from the multi-differential dilepton
spectra -- investigated in this work -- also dilepton angular
anisotropies should provide additional information for disentangling
the production channels experimentally. As analyzed in
Ref.~\cite{BCM96an} this additional observable, however, requires a
large acceptance in the dilepton spectrometer, which is presently not
fulfilled for the existing setups, but should be addressed in
experiments with HADES.

\section*{Acknowledgments}
The authors gratefully acknowledge many helpful discussions with
A.~Drees, C.~M.~Ko,  U.~Mosel and V.~D.~Toneev.  They are especially
indepted to S.~S.~Shimanskij for fruitful suggestions throughout the
course of this analysis.

\newpage

\end{document}